\pdfoutput=1
\PassOptionsToPackage{hypertexnames=false, unicode, colorlinks=true, linkcolor=linkcolor, citecolor=linkcolor, filecolor=linkcolor, urlcolor=linkcolor, pdfusetitle}{hyperref}
\documentclass[10pt,aps,prd,amsmath,floats,floatfix, twocolumn,
superscriptaddress,nofootinbib,showpacs]{revtex4-1}

\usepackage[T1]{fontenc}
\usepackage[utf8]{inputenc}
\usepackage{lmodern}
\usepackage{orcidlink}
\usepackage{verbatim}
\usepackage[bottom]{footmisc}
\definecolor{linkcolor}{rgb}{0.0,0.3,0.5}

\usepackage[all]{hypcap}
\usepackage{graphicx}
\usepackage{xspace}
\usepackage{amssymb}
\usepackage[normalem]{ulem} 
\usepackage{bm} 
\usepackage{enumitem,amssymb}
\usepackage{lineno}

\usepackage{microtype}

\usepackage[english]{babel}
\usepackage{blindtext}

\usepackage{acro}  
\DeclareAcronym{BSD}{
  short = BSD ,
  long  = Band Sampled Data
}

\DeclareAcronym{CR}{
  short = CR ,
  long  = Critical Ratio
}

\DeclareAcronym{FFT}{
  short = FFT ,
  long  = Fast Fourier Transform
}

\DeclareAcronym{CW}{
    short = CW ,
    long= continuous gravitational wave
}

\DeclareAcronym{GW}{
  short = GW ,
  long  = gravitational wave
}

\DeclareAcronym{BH}{
  short = BH ,
  long  = black hole
}

\DeclareAcronym{BSB}{
  short = BSB ,
  long  = Binary System Barycenter
}

\DeclareAcronym{SSB}{
  short = SSB ,
  long  = Solar-System Barycenter
}

\DeclareAcronym{O3}{
  short = O3 ,
  long  = the third observing run
}

\DeclareAcronym{O4a}{
  short = O4a ,
  long  = the first part of the fourth observing run
}

\DeclareAcronym{O4}{
  short = O4 ,
  long  = the fourth observing run
}

\DeclareAcronym{VBC}{
  short = VBC ,
  long  = ultralight vector boson cloud
}

\DeclareAcronym{LVK}{
  short = LVK ,
  long  = {LIGO Scientific Collaboration, Virgo Collaboration, and KAGRA Collaboration}
}

\DeclareAcronym{ASD}{
  short = ASD ,
  long  = {Amplitude Spectral Density}
}

\renewcommand{\[}{\begin{equation}}
\renewcommand{\]}{\end{equation}}
\newcommand{\med}[1]{{\rm median}\qty(#1)}

\renewcommand{\ap}{a_{\rm p}}
\newcommand{\tasc}{t_{\rm asc}}
\newcommand{\CR}{{\rm CR}}
\newcommand{\tcoh}{T_{\rm coh}}

\usepackage[arrowdel]{physics}  

\graphicspath{%
  {figs/}%
}

\DeclareMathAlphabet{\mathpzc}{OT1}{pzc}{m}{it}

\newif\ifprintauthors
\printauthorstrue  

\newcommand{\blue}{\textcolor{black}}

\newcommand{\response}{\textcolor{black}}
\newcommand{\new}{\textcolor{black}}

\newlist{todolist}{itemize}{2}
\setlist[todolist]{label=$\square$}

\begin{document}

\title{Directed searches for gravitational waves from ultralight vector boson clouds around merger remnant and galactic black holes during the first part of the fourth LIGO--Virgo--KAGRA observing run}

\ifprintauthors
\author{A.~G.~Abac\,\orcidlink{0000-0003-4786-2698}}
\affiliation{Max Planck Institute for Gravitational Physics (Albert Einstein Institute), D-14476 Potsdam, Germany}
\author{I.~Abouelfettouh}
\affiliation{LIGO Hanford Observatory, Richland, WA 99352, USA}
\author{F.~Acernese}
\affiliation{Dipartimento di Farmacia, Universit\`a di Salerno, I-84084 Fisciano, Salerno, Italy}
\affiliation{INFN, Sezione di Napoli, I-80126 Napoli, Italy}
\author{K.~Ackley\,\orcidlink{0000-0002-8648-0767}}
\affiliation{University of Warwick, Coventry CV4 7AL, United Kingdom}
\author{C.~Adamcewicz\,\orcidlink{0000-0001-5525-6255}}
\affiliation{OzGrav, School of Physics \& Astronomy, Monash University, Clayton 3800, Victoria, Australia}
\author{S.~Adhicary\,\orcidlink{0009-0004-2101-5428}}
\affiliation{The Pennsylvania State University, University Park, PA 16802, USA}
\author{D.~Adhikari}
\affiliation{Max Planck Institute for Gravitational Physics (Albert Einstein Institute), D-30167 Hannover, Germany}
\affiliation{Leibniz Universit\"{a}t Hannover, D-30167 Hannover, Germany}
\author{N.~Adhikari\,\orcidlink{0000-0002-4559-8427}}
\affiliation{University of Wisconsin-Milwaukee, Milwaukee, WI 53201, USA}
\author{R.~X.~Adhikari\,\orcidlink{0000-0002-5731-5076}}
\affiliation{LIGO Laboratory, California Institute of Technology, Pasadena, CA 91125, USA}
\author{V.~K.~Adkins}
\affiliation{Louisiana State University, Baton Rouge, LA 70803, USA}
% ...
\collaboration{The LIGO Scientific Collaboration, the Virgo Collaboration, and the KAGRA Collaboration}

\else
  \author{The LIGO Scientific
collaboration, the Virgo collaboration, the KAGRA collaboration}
\fi

\hypersetup{pdfauthor={NAME et al.}}



\begin{abstract}
We present the first directed searches for long-transient and continuous gravitational waves from ultralight vector boson clouds around known black holes (BHs). We use LIGO data from the first part of the fourth LIGO--Virgo--KAGRA observing run.
The searches target two distinct types of BHs and use two new semicoherent methods: hidden Markov model (HMM) tracking for the remnant BHs of the mergers GW230814\_230901 and GW231123\_135430 (referred to as GW230814 and GW231123 in this study), and a dedicated method using the Band Sampled Data (BSD) framework for the galactic BH in the Cygnus X-1 binary system. Without finding evidence of a signal from vector bosons in the data, we estimate the mass range that can be constrained. For the HMM searches targeting the remnants from GW231123 and GW230814, we disfavor vector boson masses in the ranges $[0.94, 1.08]$ and $[2.75, 3.28] \times 10^{-13}$~eV, respectively, at 30\% confidence, assuming a 1\% false alarm probability. 
Although these searches are only marginally sensitive to signals from merger remnants at relatively large distances, future observations are expected to yield more stringent constraints with high confidence.
For the BSD search targeting the BH in Cygnus X-1, we exclude vector boson masses in the range $[0.85, 1.59] \times 10^{-13}$~eV at 95\% confidence, assuming an initial BH spin larger than 0.5.
\\
\end{abstract}
\maketitle


\section{Introduction}
\label{sec:introduction}

Ultralight bosons are a class of theoretical particles proposed in certain extensions of the Standard Model, and their discovery could address numerous unresolved questions in particle physics and cosmology (e.g., the nature of dark matter~\cite{Bertone2018_1, Bertone2018_2, Oks2021, Baryakhtar2025} or the strong charge-parity-problem~\cite{Peccei_Quinn1977, Peccei_Quinn1977_2, Weinberg1978}). Theories predict different subclasses of ultralight bosons based on spin, including scalar (spin-0)~\cite{Peccei_Quinn1977, Peccei_Quinn1977_2, Weinberg1978, Arvanitaki2010, Freitas2021}, vector (spin-1)~\cite{Holdom1986, Goodsell2009, Jaeckel2010, Essig2013, Hui2017, Agrawal2020, Fabbrichesi2021}, and tensor (spin-2)~\cite{Clifton2012, Babichev2016, Babichev2016_2, Aoki2016, Aoki2018, Yusuke2023} fields.
Assuming only gravitational coupling, the superradiance mechanism provides a means by which ultralight bosons may form bound states around rotating \acp{BH} and grow into macroscopic clouds, producing quasi-monochromatic, long-duration \ac{GW} signals that may be observable by ground-based detectors~\cite{Arvanitaki2011, Yoshino2014, Yoshino2015, Arvanitaki2015, Arvanitaki2017, Brito2017, Brito2017_2, Baryakhtar2017, Chan2022, Cardoso2018, Baumann2019, Hannuksela2019, Zhang2019, East2017, East2017_2, East2018, Siemonsen2020, Brito2020}.

Ultralight vector bosons with mass $m_V$ can extract rotational energy from a \ac{BH} through the superradiance mechanism if the following condition is satisfied:
\begin{equation}    
    0 < \omega_{\blue{V}} < m\Omega_H,
    \label{eqn:superrad_condition}
\end{equation}
where $\omega_{\blue{V}} \approx m_V\,c^2 / \hbar$ is the angular frequency of the boson field, $m$ is the azimuthal quantum number, and $\Omega_H$ is the angular frequency of the \ac{BH}'s outer horizon.
Bound states can therefore grow exponentially in time with an instability growth rate that is maximized when the boson's Compton wavelength $\lambda$ is comparable to the \ac{BH}'s horizon radius $r_g$. 
For a superradiantly unstable state, the instability extracts energy and angular momentum fluxes from the \ac{BH} at the horizon into the bosonic field, with the change in mass directly related to the change in angular momentum.
If we assume the primary interaction of the ultralight bosons is gravitational, this results in compound field amplification and the growth of a macroscopic cloud that may extract as much as $\sim 10$\% of the \ac{BH}'s initial mass~\cite{Penrose1969, Press_Teukolsky1972, Zel'Dovich1971, Starobinskii1973, Detweiler1980, Bekenstein1973, Dolan2007, Arvanitaki2011, East2017_2, Herdeiro2022}. 

This exchange of energy and angular momentum continues until Eq.~\eqref{eqn:superrad_condition} is saturated ($\omega_{\blue{V}} = m \Omega_H$), or in other words, until the \ac{BH} has been spun down so much that superradiance can no longer occur. At this point, the cloud begins to deplete through \ac{GW} emission, which occurs at roughly twice the boson cloud oscillation frequency, with a small positive frequency drift as the magnitude of the bosons' binding energy to the \ac{BH} decreases~\cite{East2017_2, East2018, Baryakhtar2017, Isi2019, Siemonsen2020, Baryakhtar2021, Siemonsen2023_2, May2024}. For stellar and intermediate-mass \acp{BH} and bosons with masses within the range $\sim 10^{-14}$--$10^{-11}$~eV~\cite{Arvanitaki2015, Brito2017_2}, this emission frequency falls in the sensitive band of ground-based \ac{GW} detectors. Thus, current and future ground-based \ac{GW} detector networks~\cite{aLIGO, aVirgo, KAGRA, Jia2024, GW_observation_prospects, Punturo2010, Evans2021} offer a unique way to search for these ultralight bosons. If no detection is made, we are able to place constraints on the existence of ultralight bosons within the above mass range.

Many observational studies have already been designed and implemented to search for ultralight scalar bosons, and, in the absence of a detection, constrain their existence. Although a handful of studies have derived constraints on vector bosons as well, the vector signal morphology has only recently been accurately modeled via numerical calculations in both the relativistic and non-relativistic regimes,\footnote{In the relativistic regime, the \ac{BH} and the superradiant cloud are approximately the same size, whereas in the non-relativistic regime, the cloud is much larger than the \ac{BH}.} enabling more accurate predictions of the signal morphology and more robust constraints on the boson mass~\cite{Siemonsen2020}. 
Constraints are placed on the existence of ultralight scalar~\cite{Arvanitaki2015, Ng2021, Cardoso2018, Brito2017_2, caputo2025, Abac_2025_GW241011_241110} and vector~\cite{Baryakhtar2017, Cardoso2018, aswathi2025, Abac_2025_GW241011_241110} particles using \ac{BH} spin measurements. 
In some cases, however, \response{these constraints are limited by systematic uncertainties in the inferred \ac{BH} spins and masses, which depend on modelling assumptions and observational systematics~\cite{Mummery:2025tvq,Reynolds:2019uxi,McClintock:2013vwa}.}
Searches for \acp{CW} that target either the galactic center or the entire sky are used to constrain scalars and, based on assumptions about the \ac{BH} population (e.g., mass distribution, spins, and ages) and their impact on the formation and evolution of boson clouds, they disfavor the \new{scalar boson} mass range of $\sim 10^{-13}$--$10^{-12}$~eV~\cite{O3_all-sky_scalar_bosons, Palomba2019, O3_galactic_center, Zhu2020}; \new{Ref.~\cite{Dergachev2019} also briefly considered vector boson condensates in the near-zero frequency-derivative regime, estimating the distance reach for vector signals in the mass range of $\sim 10^{-14}$--$10^{-13}$~eV around \acp{BH} with a fixed spin.}
Using the null results from searches for a stochastic \ac{GW} background generated by a population of \acp{BH} with scalar~\cite{Tsukada2019, Yuan2022} and vector~\cite{Tsukada2021} clouds, boson masses $\sim 10^{-13}$~eV have been disfavored for both scalars and vectors, \response{based on assumptions regarding the \ac{BH} population and spin distributions.}
A directed search targeting the \ac{BH} in the X-ray binary Cygnus X-1 excludes scalars in the mass range of $\sim [0.6, 1] \times 10^{-12}$~eV, with the constraint depending on the assumed \ac{BH} age, which sets the timescale for boson cloud decay~\cite{O2_CygnusX1_scalar_bosons, Collaviti2024}. 
Although the results from some of these studies overlap with the parameter space of interest in this paper, \response{their interpretations rely on several assumptions and carry certain caveats, including uncertainties in \ac{BH} population and spin distributions, as well as \ac{BH} evolution history and estimated ages (see discussion above).} As a result, we cannot yet rule out the possibility that an ultralight boson exists within the disfavored mass ranges listed above.

In addition to indirect searches, ground-based interferometers are used as particle detectors for the direct detection of ultralight scalars in Refs.~\cite{Vermeulen2021, Gottel2024, Aiello2022} and vectors in Refs.~\cite{direct_dark_matter_LIGO_Virgo, direct_dark_matter_KAGRA}.
Looking to the future, the existence of a scalar cloud could impact the inspiral of binary \acp{BH}, and this may soon be detectable for certain boson masses and field strengths~\cite{Yang2018, Baumann2019, Choudhary2021}.

In this study, we present the results from the first directed searches for \ac{GW} signals from \acp{VBC} around selected known \acp{BH}, performed using data from \ac{O4a} of the \ac{LVK}. We use two semicoherent methods to target vector boson signals produced by two different categories of astrophysical sources: i) using methods outlined in Refs.~\cite{Jones2023, Jones2025} based on a hidden Markov model (HMM), we undertake the first ever directed searches for \acp{GW} from \acp{VBC} around compact binary merger remnant \acp{BH}~\cite{GWTC-4_intro, GWTC-4_methods, GWTC-4_results}, and ii) using a new method based on the so-called \ac{BSD} framework~\cite{Piccinni2019}, we search for \acp{GW} from a \ac{VBC} around the \ac{BH} in the known binary system Cygnus X-1. Each method is tailored to track signals from a different type of source; while merger remnants are isolated and are expected to emit signals with rapidly evolving frequencies, Cygnus X-1 is a binary system with a \ac{BH} that would emit a nearly monochromatic signal and has uncertain orbital parameters that must be accounted for. For a more in-depth discussion of these sources and their differences, see Sec.~\ref{sec:signal_parameters}.

One benefit of targeting known \acp{BH}---whether they are remnants from previous \ac{GW} observations or identified via X-ray emissions---is that more robust constraints can be derived. In contrast to the broader searches mentioned above, which rely on assumptions about the underlying \ac{BH} population, directed searches benefit from more complete information about the target \ac{BH}s. In particular, uncertainties in the \ac{BH} age and spin---which can strongly influence the resulting constraints---are reduced or eliminated when the source is well-characterized.
For young merger remnants, very few hypotheses are needed on the history and evolution of the \acp{BH} since their formation. On the other hand, targeting a galactic \ac{BH} like Cygnus X-1 is beneficial due to its proximity~\cite{Miller_Jones_CygX1}. In addition, because the system is much older~\cite{Wong_CygX1}, any signal observable during \ac{O4a} must correspond to an emission process that occurs over much longer timescales and produces a more slowly evolving signal (i.e., like a traditional CW), and hence the search can be run across the full \ac{O4a} dataset.\footnote{While merger remnants could produce similarly long-lived \ac{GW} emission if the bosons with appropriate masses exist, the growth time of the corresponding clouds would far exceed the \ac{O4a} time frame.}

The organization of the paper is as follows. In Sec.~\ref{sec:targets}, we identify the target \acp{BH} and \blue{describe the common evolution features of the expected \ac{GW} signals, along with target-specific parameters relevant to the searches.} 
In Sec.~\ref{sec:methods}, we outline the two search methods used in this study, \blue{each optimized for different signal durations. 
Sec.~\ref{sec:search} details the search configurations and setup for each method, with distinctions between merger remnants and persistent sources highlighted throughout. 
In Sec.~\ref{sec:follow-up}, we present the search results and explain the candidate follow-up procedures, again addressing each target in turn.}
Then, in Sec.~\ref{sec:constraints}, we estimate the range of vector boson masses that can be disfavored given that no signal is detected. We summarize our findings and conclude in Sec.~\ref{sec:conclusion}.
\blue{While the discussion is organized thematically to emphasize the shared methodological framework and signal properties, each source is discussed at every stage of the analysis to maintain clarity and completeness.}


\section{Target sources}
\label{sec:targets}

In this section, we give an overview of the \acp{BH} targeted by the two search pipelines, as well as their expected \ac{GW} signal parameters, which can be predicted using a combination of analytic~\cite{Baryakhtar2017, Rosa2011, Pani2012, Frolov2018, Baumann2019_2} and numerical~\cite{Cardoso2018, East2017, East2017_2, East2018, Siemonsen2020, Dolan2018} methods. We generate the signal parameters using the waveform model \texttt{SuperRad}, which models the dynamics, oscillation frequency, and \ac{GW} emission of ultralight boson clouds with high accuracy~\cite{Siemonsen2023, May2024}.


\subsection{Merger remnant black holes}
\label{sec:merger_remnants}

For the HMM search, we target two binary merger remnant \acp{BH} from \ac{O4a}~\cite{GWTC-4_results}: the remnants of the GW230814\_230901~\cite{GW230814} and GW231123\_135430~\cite{GW231123} mergers (henceforth referred to as GW230814 and GW231123, respectively). The estimated median parameters for each remnant are shown in Table~\ref{tab:BH_params}, where $M_i$ and $\chi_i$ are the mass and dimensionless spin of the \ac{BH} before superradiance has occurred, $D_{\mathrm L}$ is the luminosity distance, $\iota$ is the inclination angle, and RA and Dec are the right ascension and declination of the \ac{BH}. The remnants from these two events are chosen in particular because, for each BH, we find that the median estimated luminosity distance is less than the furthest reachable distance (see Fig.~9 in Ref.~\cite{Jones2023}) for a \ac{BH} with the same median mass and spin.
Note, because we use an earlier version of the parameter estimates (which were the most recent and accurate estimates available at the time of this study), the values shown in Table~\ref{tab:BH_params} vary slightly from those reported in Refs.~\cite{GWTC-4_results} and~\cite{GW231123}. However, these differences are minor and have no noticeable impact on the analyses or conclusions, as the searches are relatively insensitive to small deviations in the parameter estimates.\footnote{For GW230814, the posterior distributions of the BH properties from the preliminary version of the parameter estimation are consistent with those reported in Ref.~\cite{GWTC-4_results}. For GW231123, the posterior distributions show minor differences compared to those presented in Ref.~\cite{GW231123}. In particular, the distribution of $D_L$ in Ref.~\cite{GW231123} is more tightly constrained than the preliminary version used in this study; nonetheless, the two distributions remain consistent at the 90\% confidence level. Since we adopt the broader $D_L$ distribution, our results can be considered conservative.}

The parameter estimates used in this study have been generated with the \texttt{NRSur7dq4} waveform model~\cite{Varma2019}. This model has been chosen because it interpolates between numerical relativity data without making additional waveform modeling assumptions, and it typically performs well for signals from higher-mass sources like GW231123~\cite{GW231123}.
In standard astrophysical formation scenarios of binary systems, the high spin for the primary \blue{constituent} BH favored by the analysis of GW231123~\cite{GW231123} is in tension with the existence of a vector boson with mass that would give the loudest signal for the remnant BH~\cite{aswathi2025}. \response{Nonetheless, in general there is value in targeting remnants for a direct and independent search that is not subject to astrophysical uncertainties such as the age of the binary, spin modification through accretion, and systematic uncertainties in waveform inference~\cite{aswathi2025}, thereby providing a complementary and more direct probe of boson signals.}

\begin{table*}[tbh]
    \centering
    \setlength{\tabcolsep}{8pt}
    \renewcommand\arraystretch{1.4}
    \caption{Estimated median parameters for the remnant BHs from the GW230814 and GW231123 mergers and for the galactic \ac{BH} Cygnus X-1.}
    \begin{tabular}{lcccccc}
        \hline
        \hline
        Source & $M_i$~$[M_{\odot}]$ & $\chi_i$ & $D_{\mathrm L}$~[Mpc] & $\cos \iota$ & RA~[rad] & Dec~[rad] \\
        \hline
        Remnant from GW230814 & $58.9^{+1.8}_{-1.8}$\footnotemark[1] & $0.68^{+0.01}_{-0.02}$ & $301^{+171}_{-138}$ & $0.03^{+0.74}_{-0.67}$ & $3.21^{+2.62}_{-2.95}$\footnotemark[2] & $0.04^{+1.02}_{-1.08}$\footnotemark[2] \\
        Remnant from GW231123 & $219.8^{+22.6}_{-46.2}$\footnotemark[1] & $0.85^{+0.05}_{-0.18}$ & $2054^{+2960}_{-1280}$ & $0.45^{+0.45}_{-1.23}$ & $3.37^{+1.67}_{-0.59}$\footnotemark[2] & $0.38^{+0.40}_{-0.63}$\footnotemark[2] \\
        BH in Cygnus X-1 & \( 21.2^{+2.2}_{-2.3} \) & \( 0.95 \)\footnotemark[4] & \( 0.00222^{+0.00018}_{-0.00017}\) & \(0.887^{+0.005}_{-0.006}\) & \( 5.22883712 \)\footnotemark[3] & \( 0.61438355 \)\footnotemark[3] \\
        \hline
        \hline
    \end{tabular}
    \footnotetext[1]{We use $M_i$ and $\chi_i$ to represent the mass and dimensionless spin of the \ac{BH} before superradiance occurs, i.e., the final BH mass ($M_f$) and spin ($\chi_f$) used in the compact binary coalescence parameter estimation. The reason we do not use $M_f$ and $\chi_f$ here is to avoid confusion with the final \ac{BH} mass and spin after superradiance occurs.}
    \footnotetext[2]{See Fig.~\ref{fig:sky_maps}.}
    \footnotetext[3]{At reference epoch MJD 56198.}
    \footnotetext[4]{The spin of Cygnus X-1 is debated in the literature. In our analysis, we treat the final spin of the \ac{BH} as a free parameter and assume a nominal initial spin of \(\chi_i = 0.95\). The impact of the initial spin on the derived constraints is discussed in Sec.~\ref{sec:constraints}.}
    \label{tab:BH_params}
\end{table*}


\subsection{Cygnus X-1}
\label{sec:galactic_black_holes}

\begin{table}[hbt!]
    \caption{\label{tab:cygx1-para}Orbital parameters for Cygnus X-1.}
    \centering
    \begin{tabular}{ l c c c  }
    \hline
    \hline
    Parameter                   & Symbol       & Value                             & Ref. \\
    \hline
    Orbital Period [days]       & \( P \)      & \( 5.599829(16) \)                &~\cite{Brocksopp_CygX1} \\
    Proj. Semi-major Axis [s]\footnotemark[1]  & \( \ap \)    & \( 36.88^{+4.02}_{-3.65} \)       &~\cite{Miller_Jones_CygX1} \\
    Eccentricity [-]            & \( e \)      & \( 0.0188^{+0.0028}_{-0.0026} \)  &~\cite{Miller_Jones_CygX1} \\
    Arg. of periastron [deg]    & \( \omega \) & \( 306.6^{+6.6}_{-6.3} \)         &~\cite{Miller_Jones_CygX1} \\
    \hline
    \hline
    \end{tabular}
    \footnotetext[1]{\blue{We define the projected semi-major axis as the delay \(\ap = a\sin(\iota)/c\), with \(a\) the semi-major axis, \(\iota\) the inclination angle, and \( c \) the speed of light (\(\ap\) is given in units of light-seconds).}}
\end{table}

In the \ac{BSD}-based search, we target the \ac{BH} in the Cygnus X-1 system, with parameters listed in Table~\ref{tab:BH_params} taken from a recent study~\cite{Miller_Jones_CygX1}. Cygnus X-1 is a binary system with an orbital period \( P = 5.599829 \pm 0.000016 \) days~\cite{Brocksopp_CygX1}. Several of its orbital parameters are constrained through X-ray observations and are summarized in Table~\ref{tab:cygx1-para}. \blue{In particular, the projected semi-major axis is defined as \(\ap = a\sin(\iota)/c\), with \(a\) the semi-major axis, \(\iota\) the inclination angle, and \( c \) the speed of light,%
\footnote{\blue{Following the considerations described in Ref.~\cite{Miller_Jones_CygX1}, we assume that the \ac{BH} spin axis is aligned with the orbital angular momentum. We therefore do not distinguish between the orbital and \ac{BH} spin inclination.}}%
\,which represents the time needed for the wave to cross the system along the line of sight.} The age of the \ac{BH} is estimated to be \( 6.2 \pm 1.8 \times 10^6\) yrs~\cite{Wong_CygX1}. 

\blue{The signal's frequency varies in time due to the change of the binding energy of the cloud to the \ac{BH} as it depletes. Thus, the frequency drift happens on the same timescale as the cloud annihilation, \(\tau_{\rm GW}\). This timescale also represents the decay of the signal amplitude over time, and a detectable signal should therefore evolve over a timescale comparable to the \ac{BH}'s age, \(\tau_{\rm GW} \lesssim t_{\rm age}\).}
Given the system's age, any detectable signal is therefore expected to exhibit high stability \blue{over the observation period}, \(\mathcal{O}\qty(1\,{\rm yr})\), with only small variations in the emission frequency over time. This expectation is supported by simulations using the \texttt{SuperRad} waveform model, showing that the signal frequency drift remains within $\mathcal{O}(10^{-12})~$Hz\,s$^{-1}$ across the entire parameter space, which is well below the frequency resolution of the analysis. Consequently, we assume a monochromatic emission for the whole duration of \ac{O4a}, and we neglect frequency derivatives typically included in standard Taylor expansions (e.g.,~\cite{O3_IsolatedNS, Pia_method}).

In the literature, several studies report an extreme spin ($\geq 0.95$) for the \ac{BH} in Cygnus X-1~\cite{Gou_2011, Zhao_2021, Gou_2014, Axelsson_2011, Walton_2016}. These results disfavor the existence of a \ac{VBC} since a significant part of the rotational energy of the \ac{BH} would have been extracted through superradiance, so such a high spin could not be measured after superradiance took place. However, these measurements are impacted by systematic uncertainties and may depend on the accretion model. This is reflected by a disagreement in the literature with other studies allowing spin values between 0.5 and 0.9 ~\cite{Duro_2016, Zdziarski_2024b, Kawano_2017, Tomsick_2014, Krawczynski_2022, Krawczynski_2018}, some even compatible with a spin below 0.2~\cite{Miller_2009, Zdziarski_2024a}. In the Cygnus X-1 search, we ignore this tension by treating the measurable final spin of the \ac{BH} after superradiance as a free parameter. To estimate the \ac{VBC} properties, we adopt an initial spin value \(\chi_i = 0.95\), as shown in Table~\ref{tab:BH_params}. In Sec.~\ref{sec:constraints}, we derive constraints on the possible boson mass for different assumed values of \(\chi_i\).


\subsection{GW signal parameters}
\label{sec:signal_parameters}

Using the \texttt{SuperRad} waveform model \cite{Siemonsen2023, May2024} and the values for $M_i$ and $\chi_i$ shown in Table~\ref{tab:BH_params}, we find the optimal vector mass---that is, the mass $m_V^{\rm opt}$ that optimizes superradiant instability for a given \ac{BH}, producing the maximum \ac{GW} strain amplitude attainable by the system at a reference epoch \(t_{\rm ref}\). For the remnants from GW230814 and GW231123, \(t_{\rm ref}\) is fixed at the time when the \ac{VBC} reaches its full size~\cite{Jones2023}, corresponding to the time \(t_{\rm sat}\) when Eq.~\eqref{eqn:superrad_condition} is saturated. In the case of Cygnus X-1, we consider the start of the \ac{O4a} period as the reference time. The values of $m_V^{\rm opt}$ are shown for each source in the second column of Table~\ref{tab:signal_params}. Then, based on these values and the median BH parameters, we use \texttt{SuperRad} (for the fastest growing, initially dominant mode $m=1$) to estimate the \ac{GW} signal parameters for each BH. The parameters, also listed in Table~\ref{tab:signal_params}, are defined as follows. The values \blue{$\tau_{\rm growth}^{\rm det}$ and $\tau_{\rm GW}^{\rm det}$} are the \ac{VBC} growth and depletion timescales \blue{in the detector frame}. \blue{The parameter $\tau_{\rm growth}^{\rm det}$ represents the timescale over which the \ac{VBC} grows to its full size, and it is the inverse of the growth rate (see Eq.~4 in Ref.~\cite{Jones2023}). The parameter $\tau_{\rm GW}^{\rm det}$ is the GW emission timescale (i.e., the time over which the signal amplitude decreases by a factor of two), and it depends on the mass of the \ac{VBC} and the radiated power at saturation.}
\blue{The parameters $h_0^{\rm eff}$, $f_{\rm GW}^{\rm det}$, and $\dot{f}_{\rm GW}^{\rm det}$} are, respectively, the \ac{GW} strain amplitude \blue{(i.e., $h_0$ scaled by the orientation angle $\iota$ of the \ac{BH} as shown: $h_0^{\rm eff} = h_0\,2^{-1/2} \{[(1+\cos^2{\iota})/2]^2 + \cos^2{\iota}\}^{1/2}$~\cite{Sun2018})}, the \ac{GW} emission frequency, and the frequency drift, each evaluated in the detector frame at the reference time \((t=t_{\rm ref})\). \blue{We use the following definitions to simplify our notation: $f_{\rm GW}^{\rm det} (t_{\rm ref}) \equiv f_0$ and $\dot{f}_{\rm GW}^{\rm det} (t_{\rm ref}) \equiv \dot{f}_0$.}

\blue{Since vector bound states carry spin angular momentum, they can still grow via superradiance even with zero orbital angular momentum number ($\ell=0$). As a result, vector bound states can be concentrated closer to the \ac{BH}, leading to clouds that grow and deplete more rapidly and radiate at much higher power than their scalar counterparts~\cite{East2017, Baryakhtar2017, East2018, Siemonsen2020, Brito2020}.}
Here, we provide a depiction of how the signal frequency evolves once the VBC is saturated, independent of the specific target being searched.
At $t_{\rm sat}$, the \ac{VBC} is oscillating around the \ac{BH} with angular frequency $\omega_{\blue{V}} = m \Omega_H$~\cite{East2018}. Then, the initial frequency of the \ac{GW} emission in the source frame depends on this angular frequency: \blue{$f_{\rm GW} (t_{\rm sat}) = \omega_{V}/\pi$}.
The evolution of $f_{\rm GW}$ \blue{in the source frame} can be approximated as~\cite{May2024} 
\begin{equation}
    f_{\rm GW}(t) = f_{\infty} - \frac{|f_{\rm shift}|}{1+(t-t_{\rm sat})/\tau_{\rm GW}},
\label{eqn:GW_frequency_analytic}
\end{equation}
where \blue{$\tau_{\rm GW}$ is the source frame GW emission timescale,} $f_{\infty}$ is the asymptotic frequency at late times ($t-t_{\rm sat} \gg \tau_{\rm GW}$) and $f_{\rm shift}$ is the negative shift of $f_{\rm GW}$ away from $f_{\infty}$ as a result of the self-gravity of the \ac{VBC} at $t_{\rm sat}$~\cite{Siemonsen2023, Jones2023, May2024}, i.e., we have \blue{$f_{\rm GW} (t_{\rm sat}) = f_{\infty} - |f_{\rm shift}|$}.
While Eq.~\eqref{eqn:GW_frequency_analytic} provides a useful illustration of how the emission frequency evolves over time, the frequencies and their time derivatives shown in Table~\ref{tab:signal_params} are computed with greater accuracy using \texttt{Superrad}, which incorporates higher-order relativistic corrections~\cite{Siemonsen2023}.
Note that the signal characteristics differ between the merger remnants and Cygnus X-1, as the depletion timescales $\tau_{\rm GW}$ considered are different by orders of magnitude.

\begin{table*}[tbh]
    \centering
    \setlength{\tabcolsep}{8pt}
    \renewcommand\arraystretch{1.2}
    \caption{Estimated GW signal parameters for the remnant BHs from the GW230814 and GW231123 mergers and the BH in Cygnus X-1 using the median BH parameters and their respective $m_V^{\rm opt}$ values.}
    \begin{tabular}{lcccccc}
        \hline
        \hline
        Source & $m_V^{\rm opt}$~[10$^{-13}$~eV] & $\tau_{\rm growth}^{\blue{\rm det}}$ & $\tau_{\rm GW}^{\blue{\rm det}}$ & $h_0^{\rm eff}$ [$10^{-24}$] & $f_0$~[Hz] & $\dot{f}_0$~[Hz\,s$^{-1}$] \\
        \hline
        Remnant from GW230814 & 3.805 & 7.8~h & 28~h & 0.341 & 170.4 & $6.2 \times 10^{-7}$ \\
        Remnant from GW231123 & 1.652 & 2.3~h & 2.5~h & 1.69 & 51.2 & $9.4 \times 10^{-6}$ \\
        BH in Cygnus X-1 & 1.040 & 560~y & $7.5 \times 10^6$~y & 1.75 & 50.3 & $1.6 \times 10^{-19}$ \\
        \hline
        \hline
    \end{tabular}
    \label{tab:signal_params}
\end{table*}

As is clear from Tables~\ref{tab:BH_params} and~\ref{tab:signal_params}, the characteristics of these target \acp{BH} are quite different.
While the merger remnants are young, isolated \acp{BH} that are expected to emit short-duration, more rapidly evolving signals, Cygnus X-1 is a binary system with a much older \ac{BH} that is expected to emit a long-duration, approximately monochromatic signal. Hence, we use two different search methods, each designed to exploit a different aspect of \ac{VBC} signals: HMM is a flexible, less model-dependent method that is well-suited for signals with larger frequency drifts and greater uncertainties (see Sec.~\ref{sec:methods_HMM}), whereas the Binary BSD-VBC search technique is designed for signals with negligible frequency drifts from sources whose orbital motion must be taken into account (see Sec.~\ref{sec:methods_BSD}).

In the case of \acp{BH} from merger remnants with $D_\mathrm{L} \sim \mathcal{O}$(Gpc), we take into consideration the non-negligible impact of redshift on the \ac{GW} emission frequency and timescale.
The frequency in the detector frame scales as \blue{\mbox{$f_{\rm GW}^{\rm det} = f_{\rm GW}\,(1+z)^{-1}$}, and correspondingly, the frequency derivative in the detector frame scales as \mbox{$\dot{f}_{\rm GW}^{\rm det} = \dot{f}_{\rm GW}\,(1+z)^{-2}$} (where $\dot{f}_{\rm GW}$ is the frequency drift in the source frame).}
In a similar vein, the \ac{GW} emission timescale is modified as
\mbox{$\tau_{\rm GW}^{\rm det} = \tau_{\rm GW}\,(1+z)$}. For an in-depth discussion of how these redshift corrections affect the searches, see Sec.~IV in Ref.~\cite{Jones2023}.


\section{Methods}
\label{sec:methods}

In this paper, we use two semicoherent search methods: the HMM tracking method (Sec.~\ref{sec:methods_HMM}) and the Binary BSD-VBC method (Sec.~\ref{sec:methods_BSD}). Although both methods are based on previous work on traditional \ac{CW} searches targeting individual spinning neutron stars with nonaxisymmetries~\cite{Suvorova2016, Sun2018, Piccinni2019, O3_galactic_center}, this section describes how they have been tailored to searches for vector boson signals.

\subsection{HMM method: Merger remnants}
\label{sec:methods_HMM}

The HMM tracking technique is useful for detecting signals with wandering frequencies. It models the frequency evolution probabilistically as a Markov chain of transitions between discrete, unobservable (``hidden'') frequency states over a number of discrete time steps, and it connects these hidden states with observable data through an emission probability.
HMM tracking has been used in various searches for continuous or long-transient \ac{GW} signals~\cite{Suvorova2016, Sun2018, Isi2019, Sun2019, O2_CygnusX1_scalar_bosons, O3_supernova_remnants, ScoX1ViterbiO1, ScoX1ViterbiO2, O2SNR-Viterbi, Jones2021, Beniwal2021, O3amxp, Knee2024}.
This method is beneficial for two main reasons: i) it is computationally efficient, which is important for this study because vector boson signals live in a large, multi-dimensional parameter space, and ii) it is more capable of accounting for both the detector noise fluctuations and any uncertainties that may be present in the predicted signal waveform than other, more model-dependent \ac{CW} semicoherent search techniques (see Ref.~\cite{Riles2023}).

Reference~\cite{Isi2019} was the first to propose using an HMM-based method for follow-up searches targeting slowly-evolving scalar boson clouds around \ac{BH} merger remnants.
Building upon the work of Isi et al.~\cite{Isi2019}, Ref.~\cite{Jones2023} implemented a modified HMM search pipeline capable of tracking signals from \acp{VBC} that occur over shorter timescales \blue{[$\sim \mathcal{O}$(days-months)], making it well-suited for targets such as newly formed merger remnants.} The pipeline is an efficient semicoherent search method combining the HMM tracking scheme with a frequency-domain matched filter ($\mathcal{F}$-statistic), which quantifies the likelihood that a signal, parameterized by its frequency and associated time derivatives, is present in the data~\cite{JKS}. Previous studies have shown that this pipeline is sensitive enough to potentially detect signals from \acp{VBC} using data from current-generation detectors~\cite{Jones2023}.

In the following two subsections, we summarize the HMM search pipeline described in Ref.~\cite{Jones2023}, as well as the process of choosing search configurations for a given system, explained in Ref.~\cite{Jones2025}. Section~\ref{sec:pipeline} is meant to give a high-level overview of the HMM algorithm; for a more detailed explanation, see Refs.~\cite{Suvorova2016, Sun2018}.

\subsubsection{Search pipeline}
\label{sec:pipeline}

The search pipeline combines the HMM tracking technique with an $\mathcal{F}$-statistic, computed over discrete time segments. The $\mathcal{F}$-statistic takes as its input short Fourier transforms (SFTs) of the relevant time series data collected by each \ac{GW} detector~\cite{Cutler2005}. Note that for the simulated data used in various places throughout the study, we create these SFTs by injecting synthetic signal waveforms generated by the waveform model \texttt{SuperRad} into Gaussian noise using the simulateCW Python module in the LALPulsar library of LALSuite~\cite{lalsuite, swiglal}.

For this pipeline, we compute the $\mathcal{F}$-statistic coherently over time segments of length $T_{\rm coh}$. These coherent segments are then incoherently combined using an HMM, which is solved via the Viterbi algorithm~\cite{Viterbi1967}. The Viterbi algorithm efficiently finds the most probable path of the signal evolution (also known as the Viterbi path) in the frequency-time plane with $N_Q$ frequency bins of width $\delta f$ and $N_T$ time segments of length $T_{\rm coh}$ (see detailed formulation and descriptions in, e.g., Refs.~\cite{Suvorova2016, Sun2018}). The value $T_{\rm coh}$ (and correspondingly $\delta f$) are bounded by the predicted signal's maximum frequency derivative $\dot{f}_{\rm max}$ such that i) the signal may be thought of as monochromatic across a single time segment and ii) the signal can increase at most one frequency bin between each time segment.
The frequency bin width $\delta f$ is fixed to $1/(2T_{\rm coh})$.
Given that the maximum spin-up of the signal across the full search duration $T_{\rm obs} = T_{\rm coh} N_T$ must satisfy $\dot{f}_{\rm max} T_{\rm coh} \leq \delta f$, we have the following upper bound on $T_{\rm coh}$:
\begin{equation}
    T_{\rm coh} \leq (2 \dot{f}_{\rm max})^{-1/2}.
    \label{eqn:Tcoh}
\end{equation}
Because increasing $T_{\rm coh}$ generally improves the search sensitivity~\cite{Sun2018}, in these searches we fix $T_{\rm coh}$ to its largest possible value to maximize sensitivity for a given source model configuration (see Section~\ref{sec:configs}). 
The frequency drift is at its maximum when the \ac{VBC} reaches saturation at a time $t_{\rm sat}$. Thus, for a given system, we compute $\dot{f}_{\rm max}$ using \texttt{SuperRad} and then set $T_{\rm coh} = (2 \dot{f}_{\rm max})^{-1/2}$.
This ensures that the signal will not evolve outside HMM's tracking capabilities.

The Viterbi path returned by the search pipeline has an associated detection statistic that quantifies its significance. There are different ways to define the detection statistic (see, for example, the Viterbi score in, e.g., Ref.~\cite{Sun2018}, or the log likelihood of the optimal path, $\mathcal{L}$, in, e.g., Ref.~\cite{Sco_X-1_O3}). For the searches in this paper, we use $\mathcal{L}$ divided by the number of coherent segments $N_T$~\cite{Jones2023}, written as
\begin{equation}
    \bar{\mathcal{L}} \equiv \mathcal{L}/N_T.
    \label{eqn:detection_statistic}
\end{equation}
\blue{We use the log likelihood-based detection statistic $\mathcal{L}$ rather than, e.g., the Viterbi score because $\mathcal{L}$ has been shown to be more reliable for short duration searches~\cite{O2SNR-Viterbi, Beniwal2021, O3_supernova_remnants, O3amxp, Sco_X-1_O3}, whereas the Viterbi score is only reliable in the regime $N_Q \gg N_T$, appropriate for very long-duration searches~\cite{O2_CygnusX1_scalar_bosons, Sun2018, Isi2019}. In addition, normalizing $\mathcal{L}$ by $N_T$ makes it applicable across the many different search configurations used in this analysis.}

\subsubsection{Configurations}
\label{sec:configs}

Given a target \ac{BH}, we start by finding $m_V^{\rm opt}$. We then define a range of vector masses the pipeline \textit{may} be sensitive to, $m_V \in [0.6, 1.1]m_V^{\rm opt}$,\footnote{As shown in Ref.~\cite{Jones2023}, these searches are in fact more sensitive to some sub-optimal vector masses than the optimal mass because, for a given \ac{BH}, sub-optimal masses produce longer-lived \ac{GW} emission, allowing us to extend the $T_{\rm coh}$ used in the search.} given the \ac{BH} we consider, and we choose some number of evenly spaced $m_V$ values from within this range~\cite{Jones2025}. We must then select a set of search configurations \blue{\{$t_{\rm start}$, $T_{\rm SFT}$, $T_{\rm coh}$, $T_{\rm obs}$\}, where $t_{\rm start}$ is the GPS start time of the search (corresponding to when the \ac{VBC} has reached saturation) and $T_{\rm SFT}$ is the SFT length,} to sufficiently cover this parameter space. \blue{These configuration parameters are chosen based on the predicted signal evolution, which in turn depends on the given system parameters $\{M_i,\, \chi_i,\, D_{\rm L},\, m_V\}$.} As demonstrated in Ref.~\cite{Jones2025}, because we use the flexible HMM search technique, even a single set of search configuration parameters can recover signals generated by systems with a range of parameters. Still, we require more than one configuration to provide adequate coverage of the full parameter space. \blue{In principle, if we had unlimited computational resources, we would search using a series of optimized configurations tailored to each possible signal to maximize the sensitivity. In practice, however, we must limit the number of configurations sampled in order to balance search sensitivity with computational cost. The configurations used in this analysis are therefore chosen to span the relevant parameter space with sufficient coverage, while keeping the total cost tractable. The resulting sensitivity, accounting for the finite sampling of configurations, is quantitatively evaluated through simulations in Sec.~\ref{sec:constraints_HMM}} The process of choosing \blue{a set of} configurations is explained in detail in Appendix~A of Ref.~\cite{Jones2025} and summarized here.

We first impose several limits on the allowed search configurations to ensure computational feasibility: We require the SFT length to lie within \mbox{$15~{\rm sec} \leq T_{\rm SFT} \leq 30~{\rm min}$} (they need not be fixed to the standard $T_{\rm SFT} = 30$~min used in most \ac{CW} searches) and the coherent length within \mbox{$1~{\rm min} \leq T_{\rm coh} \leq 10~{\rm day}$}. The ratio $T_{\rm coh}/T_{\rm SFT}$ must be an integer value of at least four; that is, each detector must contribute a minimum of four SFTs per $T_{\rm coh}$ segment. Finally, we set $T_{\rm obs} = \tau_{\rm GW}^{\blue{\rm det}}$ but require that it must not exceed 180~days to minimize the computational cost of the search \blue{for sub-optimally matched boson masses}~\cite{Jones2023}.
The search computing cost for an \blue{average} system scales \blue{roughly} with the duration of the signal, ranging from approximately \blue{$\mathcal{O}$(minutes)--$\mathcal{O}$(hours)} on a single-core computer.
\blue{To cover the full parameter space for a given remnant \ac{BH} (i.e., a $\sim$ 50--100~Hz bandwidth using 11 different search configurations), the total cost of the search is approximately $\mathcal{O}(10^2)$ core-hours.}
Deriving the final sensitivity across the full parameter space via simulations (using the same configurations) typically takes $\mathcal{O}(10^4)$ core-hours per target.

Following the guidelines outlined in Ref.~\cite{Jones2025}, we randomly draw 200 posterior samples from the remnant \ac{BH}'s multidimensional posterior distribution. For each sample \ac{BH} and each value of $m_V$ we consider in the search, we find the optimal search configuration. We then independently draw 11 values from each of these three distributions at the following percentiles: 2, 10, 20, 30, 40, 50, 60, 70, 80, 90, and 98. This forms 11 search configurations with which we will run the search (see Tables~\ref{tab:config_params_GW230814} and~\ref{tab:config_params_GW231123} for the sets of configurations used in this work). For most systems drawn from the remnant's posterior distribution, more than one of these configurations should be able to recover the signal, making this spacing a conservative choice.

\subsection{Binary BSD-VBC method: Cygnus X-1}
\label{sec:methods_BSD}

The Binary BSD-VBC pipeline developed in this work builds on the \ac{BSD} framework~\cite{Piccinni2019}, which provides a compact and flexible format to manipulate and analyze calibrated strain data.
Thanks to its modularity, the \ac{BSD} framework has been applied to various \ac{CW} searches---both fully coherent~\cite{O3_millisecond_pulsar, O3_targeted_pulsar} and semicoherent~\cite{O2_galactic_center, O3_galactic_center, O3_all-sky_scalar_bosons, O3_supernova_remnants, direct_dark_matter_LIGO_Virgo, O2_isolated_NS, O3_IsolatedNS}---including the all-sky search for scalar boson clouds using data from the \ac{LVK}'s third observing run~\cite{O3_all-sky_scalar_bosons}.

The Binary BSD-VBC method adapts and extends these pipelines to target \ac{CW} signals emitted by \acp{VBC} around \acp{BH} in known binary systems. The main case study is Cygnus X-1, whose orbital parameters are well constrained by X-ray observations~\cite{Miller_Jones_CygX1, Wong_CygX1, Brocksopp_CygX1}. Details of the specific implementation for this source are given in Sec.~\ref{sec:search_BSD}.

\subsubsection{Search pipeline}
\label{sec:bsd_pipeline}

In this search, data are analyzed in sub-bands of 1~Hz overlapped by 0.5~Hz using the \ac{BSD} format. The files contain a complex time series downsampled to 1~Hz and covering the full \ac{O4a}.

The search pipeline follows a two-step approach based on standard \ac{BSD} tools. The first step aims to increase the signal coherence by coherently removing the Doppler modulation caused by the combined motion of the source and the detector. 
\blue{The data is demodulated for a chosen target coherence time, \(\tcoh\), and a given set of parameters
\(\vb\Lambda = \qty{f_0,\, {\rm RA},\, {\rm Dec},\, \ap,\, \tasc,\, \Omega,\, e,\, \omega} \), where \( f_0 \) is the emission frequency, \(\tasc\) is the time of ascending node, \( \Omega \) is the orbital angular frequency related to the orbital period \(P\) as \( \Omega = 2\pi/P \), \( e \) is the orbital eccentricity, and \(\omega\) is the argument of periapse. The correction is performed using the standard \textit{heterodyne} method implemented in the \ac{BSD} framework~\cite{Piccinni2019} and outlined in Appendix~\ref{appendix:BSD_method_heterodyne}. The heterodyne procedure produces a new time series where, for a proper choice of \(\vb\Lambda\), a signal can be considered monochromatic up to the frequency resolution \(\delta f_0 = 1/\tcoh\).}

The second step consists of a standard semicoherent search for the demodulated signal. A collection of significant peaks in the time-frequency plane, known as the \textit{peakmap}, is generated using the method detailed in Ref.~\cite{Peakmap_method}. Under proper demodulation, the signal appears in the peakmap as a line of constant frequency, with all its power contained in the same frequency bin of width \(\var f_0 = 1/\tcoh\). To identify the signal, the peakmap is projected onto the frequency axis to produce a histogram of peak counts per frequency bin. 

The significance of the number of peaks in each frequency bin is evaluated by the robust \ac{CR} statistic~\cite{Pia_method}, defined in Appendix~\ref{appendix:BSD_method_cr}. Candidates are then selected by uniformly dividing the 1~Hz band into 50 sub-bands of 0.02~Hz and choosing the frequency bin with the highest \ac{CR} statistic within each sub-band.

\subsubsection{Configurations}
\label{sec:bsd_grid}

We adopt a signal model, described in Appendix ~\ref{appendix:BSD_method_signal}, where the frequency modulation is entirely defined by a set of modulation parameters \( \vb\Lambda\). The sampling of parameters \(\vb\Lambda\) must ensure adequate coverage of the parameter space to avoid significant loss in the detection sensitivity. At the same time, the computational cost of the analysis scales with the number of templates, which must therefore be kept to a minimum. The number of templates needed to cover a parameter space is discussed in Sec.~\ref{appendix:BSD_method_grid}.

\response{The dimensionality of the parameter space to be covered can be reduced based on the considerations that we discuss below. These assumptions are not generally valid and must be validated for a given source and search setup. In particular, their validity for the Cygnus X-1 search is discussed in Sec.~\ref{sec:search_BSD}. We have verified, through injection-and-recovery of simulated signals,\footnote{\response{The templates used in this search are placed according to the binary phase metric introduced in Ref.~\cite{Paola_binary_metric}, which was derived from noise-free mismatch studies. The analysis of injected signals in simulated noise discussed in this section aims to validate the reduced phase model and the search setup for detecting signals in realistic, noisy data.}} that our search setup and simplified phase model can recover \ac{VBC} signals emitted from Cygnus X-1. The injections were performed in simulated data that mimics the detectors' noise profile.} 

The unknown signal frequency \(f_0\) can be accounted for using a correction strategy similar to the sub-band method presented in Ref.~\cite{O3_galactic_center}.
\response{Taking advantage of the BSD format, our analysis considers frequency sub-bands of 1~Hz each. In place of the unknown signal frequency, we correct the timeseries via the heterodyne procedure using the sub-band's central frequency, \(f_c\). Using an incorrect signal frequency results in a residual frequency modulation of the amplitude, \(\Delta f\), which can be neglected as long as it is fully contained within a single frequency bin, i.e., \(2\Delta f < 1/\tcoh\), thereby limiting the value of \(\tcoh\). We note that although the template frequencies are spaced by 0.5 Hz, this spacing does not affect the search frequency resolution, \(\delta f_0 = 1/\tcoh\), which is determined by the coherence time \(\tcoh\) used to create the peakmap.}

The search targets a system whose sky position is known with high precision; we therefore fix RA and Dec to the central value of their respective electromagnetic estimates. For the orbital parameters, we further assume that the angular frequency \( \Omega \), eccentricity \( e \), and argument of periastron \( \omega \) are well-constrained from electromagnetic observations, allowing the use of single central values of \( \Omega\), \(e\), and \(\omega\) for the heterodyne correction. Similar to the emission frequency, a parameter is considered to be well-constrained for a given coherence time \(\tcoh\) if, while using the central value of the parameter, the residual modulation in the heterodyned data is fully contained in a single frequency bin of the peakmap.

\response{Using the simplified model presented above,} a given 1~Hz band is analyzed by sampling a collection of heterodyne parameters \(\vb\Lambda_i = \qty{f_c,\, {\rm RA}_c,\, {\rm Dec}_c,\, a_{{\rm p}, i},\, t_{{\rm asc}, i},\, \Omega_c,\, e_c,\, \omega_c}\), where the subscript \(c\) indicates that we take a fixed central value for the parameter. The sampling of these parameters is performed to cover the two-dimensional plane \( \qty(\ap,\,\tasc) \). The time of ascending node \(\tasc\) is bounded by the orbital period as \( \tasc \in \qty[-P/2, P/2] \), and the range of \(\ap\) is constrained by electromagnetic observations of the source.
\response{The parameter-space metric of the binary phase evolution has been evaluated and tested in Ref.~\cite{Paola_binary_metric}. In particular, we consider here the short-segment regime \(\tcoh \ll P\) of the metric, which is satisfied by our search setup (see Sec.~\ref{sec:search_BSD}) and by the orbital period of Cygnus X-1 reported in Table~\ref{tab:cygx1-para}. To cover the parameter space \((\ap-\tasc)\), we use a restricted version of the binary phase metric. Following the discussion in Ref.~\cite{Paola_binary_metric}, the resolutions in \(\ap\) and \(\tasc\), while allowing for a maximal loss in signal-to-noise ratio of 10\%, are}
\[\label{eq:metric_step}
\var\ap = \frac{\sqrt{0.6}}{\pi\Omega \tcoh f_0} \qquad \var\tasc = \frac{\sqrt{0.6}}{\pi\Omega^2 \ap \tcoh f_0}.
\]
These resolutions are used to construct a square lattice \( \mathbb{Z}_2 \), as described in Refs.~\cite{Paola_binary_metric, Karl_template_placement}. Details on the grid construction can be found in Appendix~\ref{appendix:BSD_method_grid}.

Typically, an analysis satisfying these limitations would take \( < \mathcal{O}(10^5)\) core-hours to cover the reduced parameter space, such as that of Cygnus X-1. In the absence of detection, constraints can be derived very efficiently using the method outlined in Sec.~\ref{sec:bsd_upperlimits}, and the computation time needed to derive these constraints is negligible compared to the analysis time.


\subsubsection{Candidate selection and coincidences}
\label{sec:bsd_cand_selection}

For each 1~Hz band, the search is repeated for all templates \(\qty{\vb\Lambda_i}_{i=1}^N\). \blue{As presented above, for each template, we uniformly select the 50 most significant frequency bins. Combining the results across all templates, we obtain a list of candidates, from which we select a subset of the most significant ones}. Specifically, for each frequency bin of width \(\delta f_0 = 1 / \tcoh\), we select the two candidates with the highest \ac{CR} values. In this way, and for the coherence time \( \tcoh = 1000\)~s \response{used in the \ac{BSD}} search, a maximum of 2000 candidates are selected for each 1~Hz band. This step reduces the number of candidates to a manageable level. \response{We note that the selection of high \ac{CR} candidates does not impact the upper limits set by this search. Indeed, as presented in Sec.~\ref{sec:bsd_upperlimits}, the upper limits are estimated using a semi-analytic approach that considers only the loudest candidate in each 1~Hz band analyzed; weaker candidates discarded during the candidate selection are irrelevant for the upper limits estimation.}

Then, candidates are filtered to keep only those with a \ac{CR} above a threshold. Similar to Ref.~\cite{O3_supernova_remnants}, the threshold is chosen for each 1~Hz band as the mean \ac{CR} plus two standard deviations of the \ac{CR} distribution of the candidates, \blue{corresponding to a false alarm probability $P_{\rm fa} \approx 5\%$ for this first step of selection}. The \ac{CR} distribution is built by excluding candidates associated with known instrumental lines (see Sec.~\ref{sec:bsd_follow-up}). We then identify pairs of candidates coincident between the Hanford and Livingston detectors. Candidates are considered coincidental if they are on the same or adjacent frequency bins. Furthermore, coincident candidates must have compatible orbital parameters, i.e., 
\[\label{eq:metric_dist}
    d_{\rm metric} = \sqrt{ \qty( \frac{\Delta \ap}{\var \ap} )^2 + \qty( \frac{\Delta \tasc}{\var \tasc} )^2 } < 3,
\]
where \(\Delta\ap\) and \(\Delta\tasc\) denote the differences between the parameters of the candidates in each detector.

Coincident pairs of candidates, referenced hereafter as stage-2 candidates, are further analyzed in the follow-up procedure presented in Sec.~\ref{sec:bsd_follow-up}.


\section{Search setup}
\label{sec:search}

In this section, we provide details on the parameters and configurations used to run the HMM and BSD searches.
We use data taken by the two Advanced LIGO detectors, Hanford and Livingston~\cite{aLIGO}, during \ac{O4a}, which ran from 15:00 UTC on May 24, 2023 to 16:00 UTC on January 16, 2024~\cite{Capote2025, Soni2025, Ganapathy2023, Jia2024}. The HMM searches analyze data spanning two different time segments (corresponding to the weeks following both the GW230814 and GW231123 binary merger events), whereas the BSD search uses the full O4a data, for which the duty factors are \(67\%\) and \(69\%\) for Hanford and Livingston, respectively. 
All the data used in this paper, acquired when the detectors were in science observing mode, are online calibrated (i.e., low-latency C00 frames), and analysis-ready (channel names: \texttt{H1:GDS-CALIB\_STRAIN\_CLEAN\_AR} and \texttt{L1:GDS-CALIB\_STRAIN\_CLEAN\_AR})~\cite{O4cal, Wade2025, Sun2020, Viets2018, Viets_thesis, Soni2025, Vajente2020, open_data_O4a}.
Calibration uncertainties in the strain data can affect boson parameter estimates (if a signal is detected) and influence constraints or sensitivity estimates (if not). In O4a, the 1-$\sigma$ frequency- and time-dependent uncertainties are $\lesssim 10\%$ in magnitude and $\lesssim 10$~deg in phase and differ between LIGO sites~\cite{O4cal}. However, their overall impact is subdominant to noise fluctuations. We therefore do not explicitly include calibration uncertainties in our analysis.
The HMM searches take as input SFTs generated after applying a glitch gating procedure~\cite{o4gating}. The \ac{BSD} are generated from the Short FFT Database (SFDB) \cite{Peakmap_method}, and cleaned using the double-gating procedure described in Refs.~\cite{Piccinni2019, Acernese_2009}.

\subsection{HMM searches: Merger remnants}
\label{sec:search_HMM}

\subsubsection{GW230814}
\label{sec:GW230814_search}

Recalling Sec.~\ref{sec:signal_parameters} and Table~\ref{tab:signal_params}, for the remnant \ac{BH} from GW230814, we find $m_V^{\rm opt} = 3.805 \times 10^{-13}$~eV. Thus, the interesting vector mass range $[0.6, 1.1]\,m_V^{\rm opt}$ (see Sec.~\ref{sec:configs}) becomes $[2.283, 4.185] \times 10^{-13}$~eV for this search. Each boson mass within this range, if it exists, would emit \acp{GW} at a different frequency, so we have a corresponding frequency range of [97, 196]~Hz across which we run the search, divided into 1~Hz sub-bands.

\begin{figure}[hbt!]
    \centering
    \includegraphics[trim = 2cm 0cm 2cm 0cm, clip, width=1.0\linewidth]{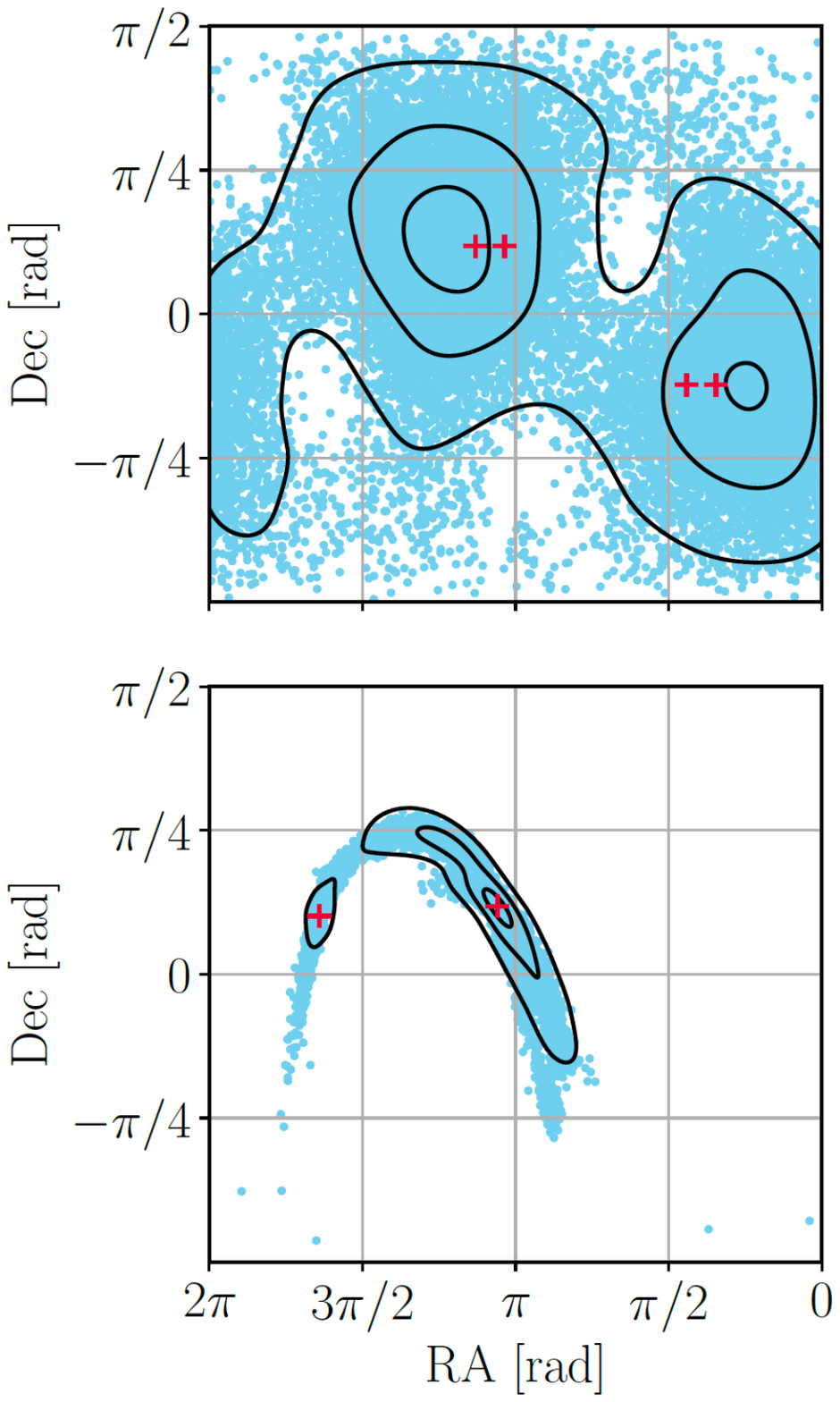}
    \caption{Joint RA and Dec posterior distribution for the merger remnants from GW230814 (top) and GW231123 (bottom). The black contours show the 90\%, 50\%, and 10\% confidence intervals. The red crosses indicate the sky positions targeted in the search for each remnant.}
    \label{fig:sky_maps}
\end{figure}

\begin{figure}[hbt!]
    \centering
    \includegraphics[width=1.0\linewidth]{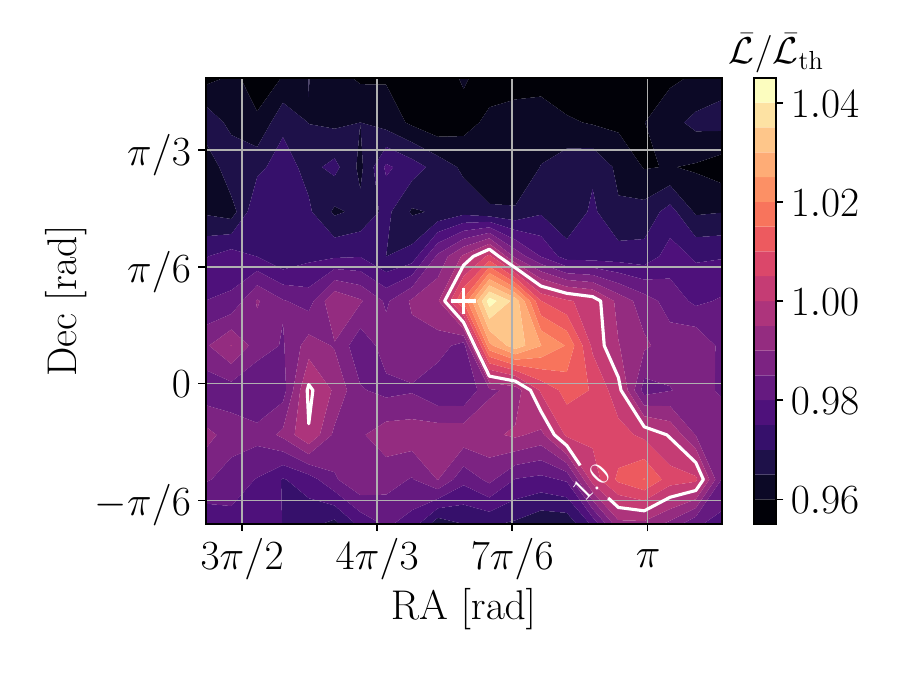}
    \caption{Colored contour of $\bar{\mathcal{L}}/\bar{\mathcal{L}}_{\rm th}$ as a function of RA and Dec for a synthetic signal injected at [3.854, 0.370]~rad, shown with a white cross marker. The signal was generated using the median \ac{BH} remnant parameters from the GW230814 merger shown in Table~\ref{tab:BH_params} and using $m_V = 2.283 \times 10^{-13}$~eV. The bright EPSF enclosed within the white contour signifies the region of the sky with $\bar{\mathcal{L}} > \bar{\mathcal{L}}_{\rm th}$ where the signal has been recovered.}
    \label{fig:GW230814_EPSF}
\end{figure}

When choosing which sky position(s) to target, it may seem reasonable to simply target the median RA and Dec listed in the table. However, because the estimated posterior distribution of the RA for this system is bimodal, the median is not representative of the data (see top panel of Fig.~\ref{fig:sky_maps}).\footnote{GW230814 was observed only by the Livingston detector, so the remnant's poorly constrained sky location aligns with the detector's antenna pattern.} Instead, we choose the optimal sky positions empirically by drawing a joint random sample of 100 RA and Dec values from the \ac{BH} posterior distribution, injecting them into Gaussian noise, and attempting to recover them using a grid of sky positions. The spacing of this grid reflects the size of the effective point spread function (EPSF) shown in Fig.~\ref{fig:GW230814_EPSF}, in which a signal has been injected at RA, Dec = [3.854, 0.370]~rad and then recovered using a grid of sky positions offset from this sky position.
Overall, the sky positions that recover the largest number of randomly sampled injections are RA, Dec = [1.088, -0.387], [1.388, -0.387], [3.254, 0.370], and [3.554, 0.370]~rad. We target all four sky positions in the search to obtain better coverage over the whole sky.\footnote{It is particularly valuable to target multiple sky positions (when the true sky position is not well constrained) for the configurations that use longer $T_{\rm coh}$ values. This is because searches with longer coherent segments are more sensitive to Doppler modulation effects and thus the assumed sky position, so there is a larger chance of missing a \ac{GW} signal if an incorrect sky position is used.} They are marked with red crosses in the top panel of Fig.~\ref{fig:sky_maps}. 
The search is not particularly sensitive to sky localization for short-duration signals with 
$\blue{T_{\rm obs}} \lesssim 1$ day, although the low sky resolution does lead to degraded sensitivity~\cite{Jones2023}. For longer signals with $\blue{T_{\rm obs}} \gtrsim 1$ day, however, a mismatch in sky position can still cause a marginal signal to be missed. This uncertainty due to poorly-constrained sky localization is incorporated into the sensitivity estimates we present in Sec.~\ref{sec:constraints_HMM}.

Using the method outlined in Appendix~A of Ref.~\cite{Jones2025} \blue{(and summarized in Sec.~\ref{sec:configs} above)} for choosing search configurations, we use 11 different configurations in the search, shown in Table~\ref{tab:config_params_GW230814}.
In particular, we choose a set of $T_{\rm coh}$ values to cover the full range of $\dot{f}_{\rm max}$ values that may occur for the system: $\sim \mathcal{O}[10^{-9}, 10^{-6}]~$Hz\,s$^{-1}$.
Across all 11 configurations, we run the search on detector data spanning the GPS times [1376111180, 1385937626]~s. Gaps are present in the available data from Hanford and Livingston during this time period, resulting from both scheduled maintenance and unexpected lock loss due to various environmental disturbances. 
In particular, although the duty factor is $\sim 70\%$ for each detector during the analysis time of GW230814, many of the data gaps lie within roughly two days post-merger, when the signal is expected to be strongest.
While the HMM search pipeline is designed to accommodate data gaps, the less data that is available, the less sensitive the search becomes. This decreased sensitivity is reflected in the eventual estimated vector mass range that is disfavored by targeting this remnant (see Sec.~\ref{sec:constraints_HMM}).

\subsubsection{GW231123}
\label{sec:GW231123_search}

As in the previous section, we refer to Table~\ref{tab:signal_params} for the optimal boson mass corresponding to the remnant \ac{BH} from GW231123: $m_V^{\rm opt} = 1.652 \times 10^{-13}$~eV. Thus, for the search we consider the range $m_V = [0.901, 1.502] \times 10^{-13}$~eV and the corresponding frequency band [21, 73]~Hz, split into 1~Hz sub-bands.

Once again, the RA posterior distribution shows some bimodality, but because the sky position is significantly more well-constrained than GW230814 [i.e., the 90\% confidence interval is constrained to $\mathcal{O}(10^3)$~deg$^2$], for the search we simply target the two local maxima marked by red crosses in the bottom panel of Fig.~\ref{fig:sky_maps}, located at RA, Dec = [3.329, 0.372] and [5.153, 0.318]~rad. These two local maxima are tested using the same method described above for GW230814, and we find that they provide sufficient coverage across the 90\% credible region of the sky position.

We start by identifying 11 potential search configurations at the same percentiles used for GW230814. However, two configurations cannot be used in the search, as their GPS start times (1602835969 and 2785340449 for the 90th and 98th percentiles, respectively) do not fall within \ac{O4}.\footnote{These times correspond to a non-optimal region of the parameter space where the \ac{VBC} grows and dissipates very slowly and thus it will be $\mathcal{O}$(years) before the \ac{GW} emission reaches its peak.}
Thus, we limit the number of configurations used in the search to only the first 9, shown in Table~\ref{tab:config_params_GW231123}.
Again, we choose the set of $T_{\rm coh}$ values to cover the potential $\dot{f}_{\rm max}$ range $\sim \mathcal{O}[10^{-15}, 10^{-5}]~$Hz\,s$^{-1}$ for this system.
Gaps are again present in the data available from both detectors, limiting the search sensitivity; while the duty factor for Hanford is 68\% across the relevant time frame, for Livingston it is only 51\%.

\begin{table}
    \centering
    \setlength{\tabcolsep}{2.5pt}
    \renewcommand\arraystretch{1.2}
    \caption{Search configuration parameters and the percentiles at which they are drawn for the BH remnant from the GW230814 merger. These percentiles, \blue{taken from the distribution of search configurations generated for 200 randomly sampled \ac{BH} posteriors $\times\,11$ values of $m_V$,} are chosen so that the configurations used in the search adequately cover the remnant's full posterior distribution.}
    \begin{tabular}{lcccc}
        \hline
        \hline
        Percentile & $T_{\rm SFT}$~[m] & $T_{\rm coh}$~[m] & $T_{\rm obs}$~[m] & GPS start time~[s] \\
        \hline
        2 & 2.85 & 11.4 & 991.8 & 1376111180 \\
        10 & 3.4 & 13.6 & 1278.4 & 1376115080 \\
        20 & 3.9 & 15.6 & 1653.6 & 1376119160 \\
        30 & 4.55 & 18.2 & 2311.4 & 1376124380 \\
        40 & 5.75 & 23.0 & 3519.0 & 1376134280 \\
        50 & 7.45 & 29.8 & 5632.2 & 1376148320 \\
        60 & 10.45 & 41.8 & 9823.0 & 1376172680 \\
        70 & 15.05 & 60.2 & 17759.0 & 1376211140 \\
        80 & 24.4 & 97.6 & 36990.4 & 1376285360 \\
        90 & 26.85 & 161.1 & 79583.4 & 1376412320 \\
        98 & 27.65 & 248.85 & 155780.1 & 1376590820 \\
        \hline
        \hline
    \end{tabular}
    \label{tab:config_params_GW230814}
\end{table}

\begin{table}
    \centering
    \setlength{\tabcolsep}{2.5pt}
    \renewcommand\arraystretch{1.2}
    \caption{Search configuration parameters and the percentiles at which they are drawn for the BH remnant from the GW231123 merger. These percentiles, \blue{taken from the distribution of search configurations generated for 200 randomly sampled \ac{BH} posteriors $\times\,11$ values of $m_V$,} are chosen so that the configurations used in the search adequately cover the remnant's full posterior distribution.}
    \begin{tabular}{lcccc}
        \hline
        \hline
        Percentile & $T_{\rm SFT}$~[m] & $T_{\rm coh}$~[m] & $T_{\rm obs}$~[m] & GPS start time~[s] \\
        \hline
        2 & 0.6 & 2.4 & 84.0 & 1384788409 \\
        10 & 1.0 & 4.0 & 172.0 & 1384792009 \\
        20 & 1.45 & 5.8 & 301.6 & 1384796449 \\
        30 & 2.25 & 9.0 & 630.0 & 1384804309 \\
        40 & 3.55 & 14.2 & 1235.4 & 1384816729 \\
        50 & 6.0 & 24.0 & 2424.0 & 1384837489 \\
        60 & 9.4 & 37.6 & 4587.2 & 1384866949 \\
        70 & 14.35 & 57.4 & 9011.8 & 1384911769 \\
        80 & 26.2 & 157.2 & 42601.2 & 1385145169 \\
        \hline
        \hline
    \end{tabular}
    \label{tab:config_params_GW231123}
\end{table}


\subsection{BSD search: Cygnus X-1}
\label{sec:search_BSD}

Using the median values for the parameters of Cygnus X-1 reported in Table~\ref{tab:BH_params}, we find the boson mass $m_V^{\rm opt} = 1.040 \times 10^{-13}$~eV produces the signal with the maximum strain amplitude at the start of \ac{O4a}. Taking into account the uncertainties in the \ac{BH} parameters, we estimate the signal amplitudes corresponding to boson masses around $m_V^{\rm opt}$ using the \texttt{SuperRad} model. Then, comparing these values to the minimum detectable strain given in Eq.~(67) of Ref.~\cite{Pia_method}, we select the frequency band \( f_0 \in [24.5, 125.5]\,\mathrm{Hz} \) as the range of frequencies where a signal could be detected.  Therefore, the search spans 202 distinct 1~Hz bands, overlapped by 50\%.

We fix the coherence time to \( \tcoh = 1000\)~s for all frequency bands investigated. This value is motivated by multiple factors. Demodulating the data using the central frequency of the band rather than the real---unknown---frequency of the signal leads to a residual modulation of the signal. Our choice of \(\tcoh\)  ensures that, for any signal present in the band, the residual modulation \(\Delta f\) is smaller than the size of a frequency bin, i.e. \mbox{\(2\Delta f < 1/\tcoh\)}. Similarly, this coherence time ensures that the uncertainties in the eccentricity parameters of Cygnus X-1's orbit, reported in Table~\ref{tab:cygx1-para}, are fully covered by a template using the central values of the parameters. At the same time, a coherence time \( \tcoh = 1000 \)~s keeps the analysis time to a realistic level with \(\mathcal{O}(60\,000)\)~core-hours being needed to analyze the parameter space of Cygnus X-1 in both LIGO detectors.

Following the grid construction method presented in Sec.~\ref{sec:bsd_grid}, for each band we define a set of parameter points \(\qty{ \vb{\Lambda}_i}_{i=1}^{N}\) covering the uncertainties in the orbital parameters of Cygnus X-1 reported in Table~\ref{tab:cygx1-para}. The time of superior conjunction \(T_0\) of Cygnus X-1 has been estimated in  Ref.~\cite{Brocksopp_CygX1}, but the extrapolation of this value to estimate the time of ascending node \(\tasc\) at the \ac{O4a} period leads to significant uncertainties compared to our search resolution. We choose an agnostic approach where the range of all possible values \(\tasc \in [-P/2, P/2]\) has been covered by the search. The number of templates per band ranges from \( N = 1\,960 \) at 24.5~Hz to \( N = 43\,416\) at 125.5~Hz, in agreement with Eq.~\eqref{eq:grid_size}. Combining the number of templates used in all the bands, the search uses 3\,687\,225 templates.

The search is performed with bands overlapping by 0.5~Hz to avoid cropping signals close to the band edges during \ac{BSD} creation. Each band is searched independently, and each produces its own set of stage-2 candidates, which are selected following the discussion in Sec.~\ref{sec:bsd_cand_selection}.


\section{Candidate follow-up}
\label{sec:follow-up}

In this section, we outline the follow-up procedures used to eliminate any search candidates whose origins are not astrophysical. See Tables~\ref{tab:HMM_vetoes} and~\ref{tab:BSD_vetoes} for the numbers of candidates that remain from the HMM and BSD searches after each veto procedure described in the following sections.


\subsection{HMM searches: Merger remnants}
\label{sec:follow-up_HMM}

For GW230814, we run the search over 99 individual 1~Hz bands, 11 search configurations, and 4 sky positions. This is 4356 iterations in total. Similarly, for GW231123, we have 52 1~Hz bands, 9 search configurations, and 2 sky positions, yielding 936 search iterations. For each 1~Hz band, we require the detection statistic $\bar{\mathcal{L}}$ to exceed a threshold $\bar{\mathcal{L}}_{\rm th}$ corresponding to a 1\% false alarm probability ($P_{\rm fa}$).\footnote{The 1\% false alarm probability corresponds to each configuration in each 1~Hz band.} The value of $\bar{\mathcal{L}}_{\rm th}$, which varies for each $T_{\rm coh}$ used in the searches, is obtained empirically as follows: For a given value of $T_{\rm coh}$, we run 300 searches in pure Gaussian noise (with \ac{ASD} = $4 \times 10^{-24}$~Hz$^{-1/2}$) at a randomly chosen 1~Hz frequency band, and we define $\bar{\mathcal{L}}_{\rm th}$ as the value of $\bar{\mathcal{L}}$ that lies at the 99th percentile of these results.

After implementing this initial threshold cut, we have 421 and 285 signal candidates across all configurations for GW230814 and GW231123, respectively. However, many candidates actually appear to be the same candidates picked up by different configurations (e.g., for GW230814, a candidate at $\sim 102.1$~Hz is identified for nearly every value of $T_{\rm coh}$ and at every sky position). We expect a large fraction of these first-pass candidates to be simply the result of noise artifacts (e.g., power line harmonics, thermally excited mirror suspension violin modes~\cite{Covas2018, Soni2025}) and non-Gaussianities in the interferometric data. We use the known-line~\cite{known_lines_O4a} and single-interferometer veto techniques described in Appendix~\ref{sec:HMM_vetoes} (which are commonly used in \ac{CW} searches) to help distinguish candidates of this nature from a true astrophysical signal~\cite{Sco_X-1_O3}. We manually inspect any candidates that survive these initial vetoes by assessing their consistency with the signal model, scrutinizing the spectrograms, identifying clear characteristics of noise (such as a candidate occurring in only one detector due to short-period artifacts and failing to meet the stringent criteria of the single-interferometer veto), etc. For a detailed description, see Appendix~\ref{sec:HMM_vetoes}.
Table~\ref{tab:HMM_vetoes} shows the candidates that remain for each merger remnant after each veto. After manual inspection, no signal candidates remain.

\begin{table}[tbh]
    \centering
    \setlength{\tabcolsep}{8pt}
    \renewcommand\arraystretch{1.2}
    \caption{Number of candidates remaining from the search targeting the remnants from the GW230814 and GW231123 mergers after each veto has been applied.}
    \begin{tabular}{lll}
        \hline
        \hline
         & \multicolumn{2}{l}{Remaining candidates} \\
        Stage & GW230814 & GW231123 \\
        \hline
        Initial candidates & \multicolumn{1}{c}{421} & \multicolumn{1}{c}{285} \\
        Known-line veto & \multicolumn{1}{c}{44} & \multicolumn{1}{c}{209} \\
        Single-interferometer veto & \multicolumn{1}{c}{19} & \multicolumn{1}{c}{101} \\
        Manual inspection & \multicolumn{1}{c}{0} & \multicolumn{1}{c}{0} \\
        \hline
        \hline
    \end{tabular}
    \label{tab:HMM_vetoes}
\end{table}


\subsection{BSD search: Cygnus X-1}
\label{sec:bsd_follow-up}

From the \ac{BSD}-based search, we identify 27 stage-2 candidates that pass the selection process described in Sec.~\ref{sec:bsd_cand_selection}. We follow up these candidates using the vetoing procedure described in Appendix~\ref{appendix:BSD_follow_up}, \blue{corresponding to an overall false alarm probability $P_{\rm fa} < 0.1\%$}. The list of candidates is first filtered by removing those associated with known instrumental lines~\cite{known_lines_O4a}, reducing the list to eight candidates. 
During the creation of candidate pairs (see Sec.~\ref{sec:bsd_cand_selection}), the same candidate could be matched to multiple coincident candidates in the other detector. To remove this redundancy, we cluster all pairs that share a common candidate, reducing the number of independent candidates to six. We then apply standard vetoes adapted from previous \ac{BSD}-based searches (e.g.,~\cite{O3_galactic_center, O3_supernova_remnants}), which are based on the consistency of the candidates with the parameters of Cygnus X-1 and with the expected behavior of the detection statistic. Three candidates pass these vetoes and are further inspected. All three are consistent with artifacts produced by non-Gaussianities in the detector. More details on the follow-up procedure are given in Appendix~\ref{appendix:BSD_follow_up}. Table~\ref{tab:BSD_vetoes} summarizes the number of candidates remaining after each veto.

\begin{table}[tbh]
    \centering
    \setlength{\tabcolsep}{8pt}
    \renewcommand\arraystretch{1.2}
    \caption{Number of candidates remaining from the Cygnus X-1 search after each veto has been applied.}
    \begin{tabular}{lc}
        \hline
        \hline
         & Remaining candidates \\
        Stage & Cygnus X-1 \\
        \hline
        Initial candidates & 27\\
        Known-line veto & 8\\
        Clustering & 6\\
        Astrophysical consistency veto & 5\\
        Statistical consistency check & 3\\
        Manual inspection & 0\\
        \hline
        \hline
    \end{tabular}
    \label{tab:BSD_vetoes}
\end{table}


\section{Constraints}
\label{sec:constraints}

In this section, we estimate the vector boson mass range that can be constrained given the absence of a confident detection in the searches described above. 


\subsection{HMM searches: Merger remnants}
\label{sec:constraints_HMM}

After applying the veto procedures described in Sec.~\ref{sec:follow-up_HMM}, all candidates from the HMM searches are eliminated.
In this section, we investigate the confidence with which we disfavor the existence of a given vector boson mass range.
We adopt an empirical approach in which synthetic signals are injected into simulated Gaussian noise configured to match the real data, with \acp{ASD} derived from detector data at the corresponding times and frequencies, and with data gaps reproduced to reflect those present during the analysis period.
We marginalize the detection probabilities over the \ac{BH} parameter uncertainties to reduce potential biases in the sensitivity estimates.
We make a common assumption when interpreting the search results: the vector field interacts only gravitationally, with no additional interactions or couplings to the Standard Model. See Sec.~IV of Ref.~\cite{Jones2025} for details on how this assumption can be partially lifted.

In Ref.~\cite{Jones2025}, a framework is developed for constraining the boson mass that marginalizes over the parameter uncertainties typical to a binary merger remnant detected gravitationally. We start by drawing a number $N_{\rm BH}$ of random samples from the \ac{BH} posterior distribution with parameters $\boldsymbol{\theta}_i$. Then, for a given boson mass, we generate a synthetic \ac{GW} signal for each sample \ac{BH} and inject the signal into a number $N_{\rm noise}$ of random Gaussian noise simulations. We evaluate the recovery rate across all sampled systems and noise realizations \blue{as shown~\cite{Jones2025}:}
\begin{equation}
    P_{\rm det}(m_V) = \frac{1}{N_{\rm BH} N_{\rm noise} } \sum_{i=1}^{N_{\rm BH}} N_{\rm det}(\boldsymbol{\theta}_i; m_V),
    \label{eqn:P_det}
\end{equation}
where $N_{\rm det}(\boldsymbol{\theta}_i; m_V)$ is the number of recovered ($\bar{\mathcal{L}} > \bar{\mathcal{L}}_{\rm th}$) signals out of $N_{\rm noise}$ noise realizations for a given vector mass $m_V$ and set of \ac{BH} parameters $\theta_i$. This value $P_{\rm det}(m_V)$ can be interpreted as the confidence to which the existence of the vector boson with mass $m_V$ can be excluded given that no signal is detected in the searches.

\begin{figure*}[hbt!]
    \centering
    \includegraphics[width=0.95\linewidth]{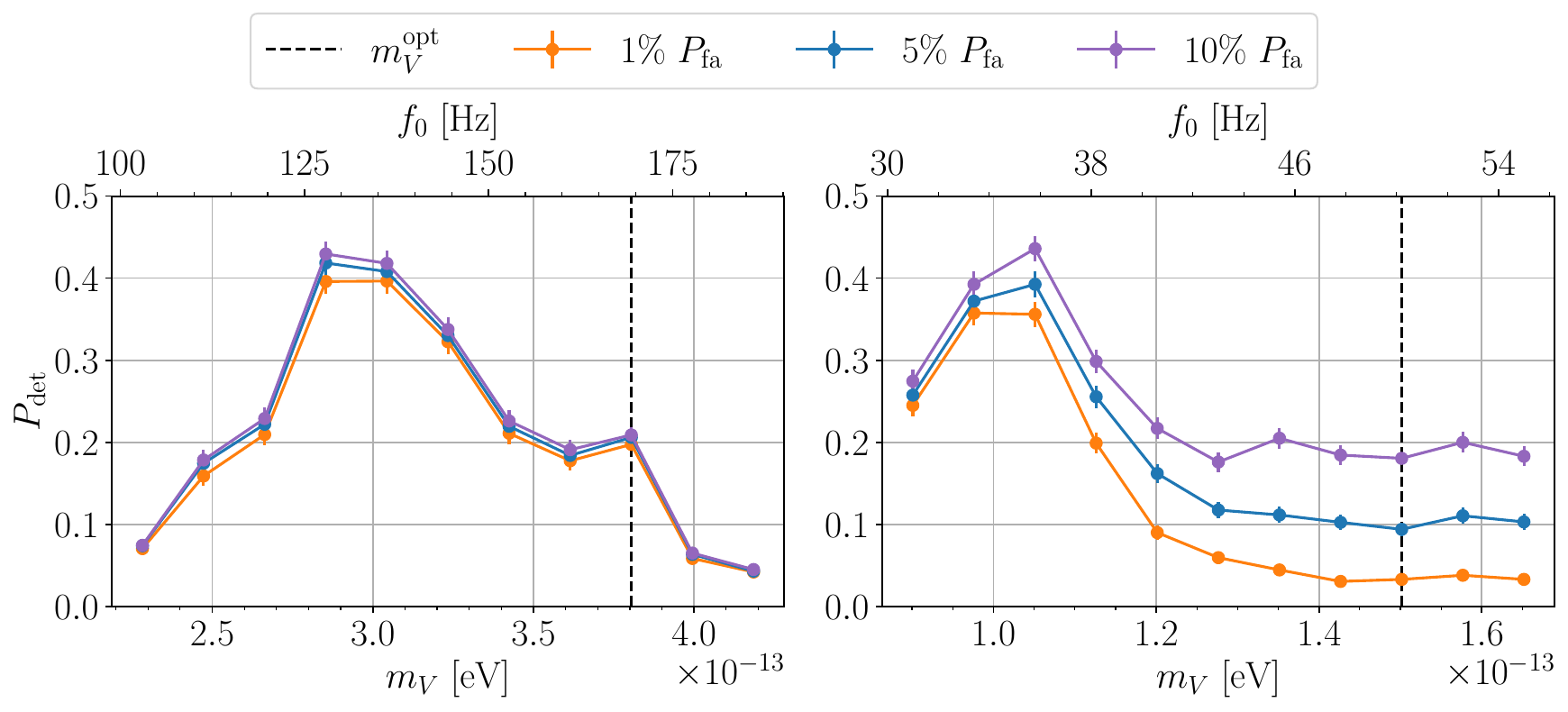}
    \caption{The detection probability $P_{\rm det}$ as a function of $m_V$ (bottom axis) and the corresponding \ac{GW} frequency $f_0$ in the detector frame (top axis) with a 1\% (orange), 5\% (blue), and 10\% (purple) false alarm probability for the remnant BHs from the GW230814 (left) and GW231123 (right) mergers. A redshift correction has been applied to obtain $f_0$ in the detector frame using the median $D_L$ value for each remnant (as reported in Table~\ref{tab:BH_params}). The vertical dashed lines mark the optimal boson masses $m_V^{\rm opt}$ for the remnants with median parameters also shown in Table~\ref{tab:BH_params}.
    The error bars represent the $1\sigma$ beta-binomial uncertainty in $P_{\rm det}$~\cite{beta-binomial}.}
    \label{fig:Pdet_v_mV}
\end{figure*}

In Fig.~\ref{fig:Pdet_v_mV} we show the confidence with which we disfavor the vector mass using this procedure for GW230814 (left panel) and GW231123 (right panel), where, following the guidelines in Ref.~\cite{Jones2025}, we have used $N_{\rm BH} = 200$ and $N_{\rm noise} = 10$. We run the simulations across all configurations used in the real searches (i.e., \{RA, Dec\} and \{$T_{\rm SFT}$, $T_{\rm coh}$, $T_{\rm obs}$, $t_{\rm start}$\}). The synthetic signal is considered recovered if at least one configuration returns an above-threshold detection statistic. The orange, blue, and purple lines in the figure indicate a 1\%, 5\%, and 10\% $P_{\rm fa}$ threshold, respectively.
$P_{\rm det}$ indicates the confidence with which one can disfavor a vector mass range given the null search results (i.e., if some range of data points lie above a given $P_{\rm det}$ value, that mass range is disfavored with $P_{\rm det}$ confidence). While the searches targeting the remnant from GW230814 and GW231123 are not sensitive enough to constrain the vector mass at high confidence, the following is an example of how we would estimate the boson mass range that can be constrained: We disfavor the vector mass ranges $[2.75, 3.28] \times 10^{-13}$ and $[0.94, 1.08] \times 10^{-13}$~eV for GW230814 and GW231123, respectively, with 30\% confidence for $P_{\rm fa} = 1\%$.

These results show that we are approaching the sensitivity required to place robust constraints on the vector boson mass. As the detectors continue to undergo improvements in subsequent observing runs, we anticipate a growing number of high-SNR events, which will enable increasingly sensitive searches and stronger constraints across a wide range of the mass parameter space. In addition, once the vector mass can be constrained with higher confidence, the results can be mapped to other interaction models~\cite{Jones2025}.


\subsection{BSD search: Cygnus X-1}
\label{sec:constraints_BSD}

Among the 27 stage-2 candidates obtained from the \ac{BSD} search, none pass the follow-up procedure. In the absence of a plausible signal candidate, we set upper limits on the strain amplitude of \acp{CW} emitted by a \ac{VBC} surrounding Cygnus X-1. In the following subsections, we outline the procedure used to estimate these upper limits and describe how they are subsequently translated into constraints on the mass of a hypothetical ultralight vector boson.

\subsubsection{Upper limits and Sensitivity}
\label{sec:bsd_upperlimits}

In this section, we estimate upper limits on the strain amplitude \( h_0 \), defined as the maximum amplitude above which the presence of a \ac{CW} signal can be excluded at a given confidence level. The limits are computed using a conservative semi-analytical method previously applied in Refs.~\cite{O3_galactic_center, O3_IsolatedNS, O3_all-sky_scalar_bosons}. The method evaluates the upper limits in each 1~Hz frequency band using the analytical relation \cite{O3_galactic_center, O3_IsolatedNS, O3_all-sky_scalar_bosons}
\[\label{eq:bsd_sensi}
    h_{\rm UL}^{95\%} \approx \frac{\mathcal{B}}{N^{1/4}} \sqrt{\frac{S_n(f)}{\tcoh}} \sqrt{ \CR_{\rm max} + 1.645},
\]
where \( N \sim T_{\rm obs}/\tcoh \) is the number of segments used to construct the peakmap, the value 1.645 is computed from Eq.(67) in Ref.~\cite{Pia_method} fixing the confidence level to 95\%, \(S_n^{1/2}(f)\) is the detector average \ac{ASD}, and \(\CR_{\rm max}\) is the maximum \ac{CR} value observed in the band. The coefficient \( \mathcal{B} \) accounts for the average detector response to the source sky position and signal polarization and depends on the peak selection threshold. For this search and the sky position of Cygnus X-1, we evaluate \( \mathcal{B} \approx 3.37 \) in Livingston and \(\mathcal{B} \approx 3.30\) in Hanford. The computation of these factors is detailed in Appendix~\ref{appendix:BSD_ul} following the discussions in Refs.~\cite{O3_galactic_center, Pia_method, peakmap_sensi_note}.
The final upper limit placed in each band is taken as the less constraining limits of the two detectors.

We have verified that this semi-analytical approach yields conservative upper limits compared to those obtained with a classical frequentist approach based on simulated signal injections, while requiring fewer computational resources. This verification was performed on ten 1~Hz frequency bands, randomly selected within the search frequency range, using simulated data that reproduces the detectors' noise levels. The validation complements previous checks carried out for other observing runs of the LIGO and Virgo detectors and for similar \ac{BSD}-based methods \cite{dicesarethesis, O3_galactic_center}.

\begin{figure}[hbt!]
    \centering
    \includegraphics[width=1\linewidth]{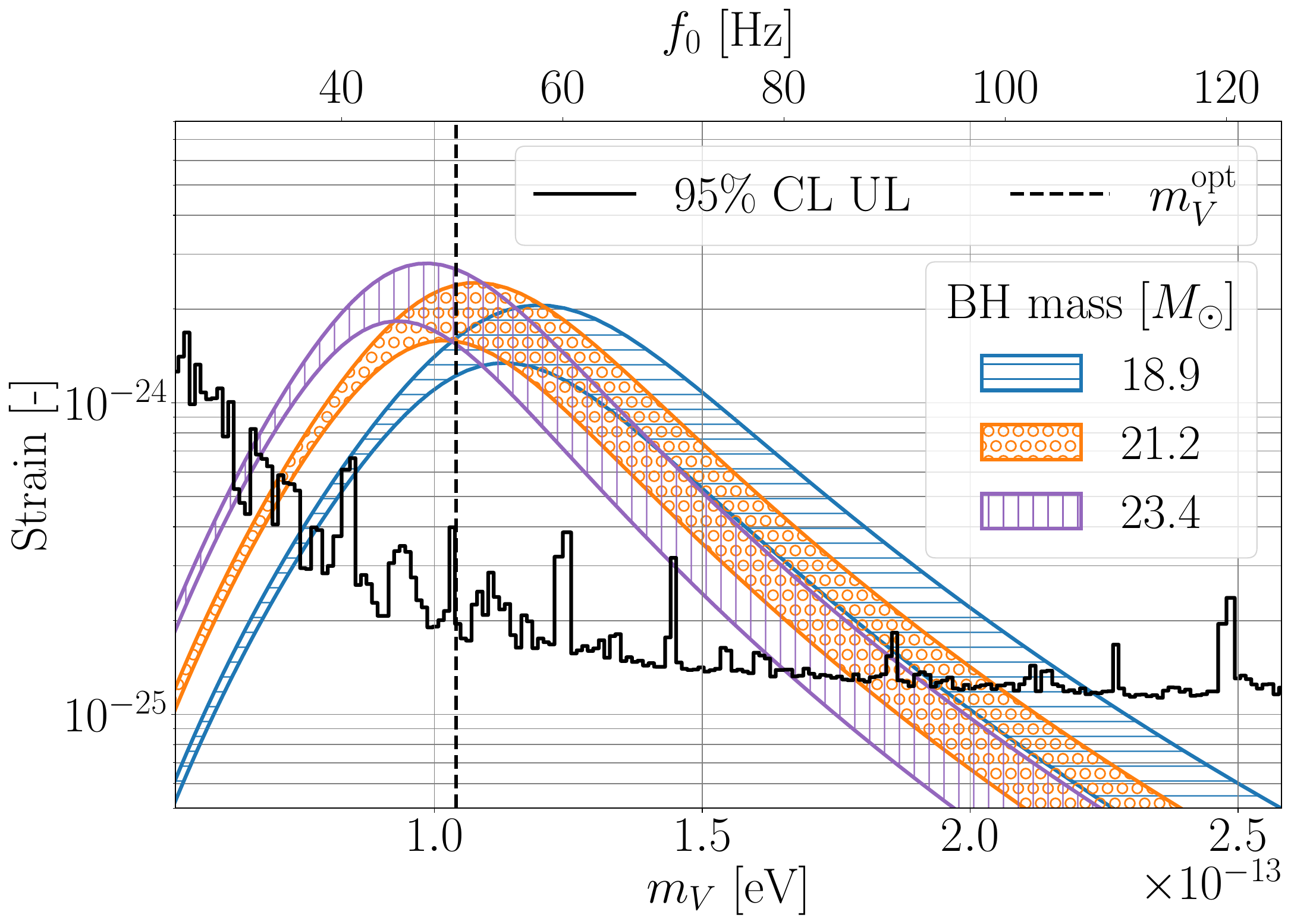}
    \caption{\label{fig:bsd_ul} Upper limit estimates at a 95\% confidence level (black curve) as a function of frequency (top axis) and the corresponding boson mass assuming the central values for the mass and age of Cygnus X-1 (bottom axis). The hashed regions correspond to the predicted strain amplitude from a \ac{VBC} around Cygnus X-1, assuming the central value (orange, circles), 5th percentile (blue, horizontal lines), and 95th percentile (purple, vertical lines) of the  \ac{BH}  mass posterior distribution from~\cite{Miller_Jones_CygX1}. We assume the \ac{BH} initial spin to be \(\chi_i = 0.95\) and the hashed regions account for uncertainties in the \ac{BH} age and distance.}
\end{figure}

In Fig.~\ref{fig:bsd_ul}, we show the joint upper limits at a 95\% confidence level, reporting in each frequency band the worst-case estimate between the two detectors. We also report the boson masses corresponding to the frequency axis, assuming the central astrophysical parameters of Cygnus X-1 reported in Table~\ref{tab:BH_params}.

We quantify the performance of the search using the \textit{sensitivity depth}~\cite{Behnke2015}, defined as
\[
    \mathcal{D}^{95\%} = \frac{\sqrt{S_n(f)}}{h_{\rm UL}^{95\%}}.
\]
This quantity has become a key figure for assessing the performance of a \ac{CW} search configuration, independent of the noise level~\cite{O3a_BinaryNS, O3_IsolatedNS, O3_rmodes, O3a_cassiopeia_velajr}. For the BSD search, we report a sensitivity depth of \(37.22\,\rm{Hz}^{-1/2}\) in the band centered on \(50.5~\rm{Hz}\) corresponding to the optimal boson mass for Cygnus X-1 (see Table~\ref{tab:signal_params}). The average depth over all the analyzed bands is \(35.15\,\rm{Hz}^{-1/2}\) and is more or less constant in all the bands, excluding the one with significant noise disturbances.

\subsubsection{Constraints}
\label{sec:bsd_implications}

The upper limits obtained in the previous section can be used to constrain the existence of ultralight vector bosons by comparing them to the expected strain amplitude of a signal emitted by a \ac{VBC} in Cygnus X-1. The amplitude is estimated assuming the central value, the 5th, and the 95th percentiles of the \ac{BH} mass posterior distribution reported in Ref.~\cite{Miller_Jones_CygX1}. We also consider the uncertainties in the \ac{BH} age and distance and assume the \ac{BH} spin before the superradiant instability to be \(\chi_i = 0.95\). The amplitude is then evaluated at different boson masses (i.e., at different frequencies) using the \texttt{SuperRad} model~\cite{Siemonsen2023, May2024}. The simulated strain amplitudes are shown in Fig.~\ref{fig:bsd_ul}, along with the upper limits derived in the previous section.

By comparing the upper limits with the expected strain amplitude, we can exclude the presence of a signal emitted by a \ac{VBC} in Cygnus X-1 in every frequency band in the range 41.25--80~Hz, with the exception of the narrow band 69.75--70.25~Hz. The absence of a signal in these bands excludes the existence of an ultralight vector boson with a mass in the range $[0.85, 1.65] \times 10^{-13}$~eV (excluding the band $[1.44, 1.45] \times 10^{-13}$~eV).

These exclusion regions are obtained under the assumption that the initial spin of the \ac{BH} was \(\chi_i = 0.95\). This assumption can be relaxed by computing the exclusion region for different initial spin values. As the assumed initial spin decreases, the expected amplitude decreases accordingly. Nevertheless, based on \texttt{SuperRad} simulations, for any value \(\chi_i > 0.5\), our upper limits can constrain the existence of a signal between \(41.25\) and \(77.25\)~Hz, corresponding to boson masses within the range $[0.85, 1.59] \times  10^{-13}$~eV (with the same excluded band as before). For spin values \(\chi_i < 0.5\), the lower expected amplitude narrows down the constrained frequency range. At \(\chi_i = 0.2\), we are still able to constrain frequencies in the range 50.25--77.25~Hz and boson masses $[1.03, 1.59] \times 10^{-13}$~eV. 
\blue{For lower initial spins \(\chi_i < 0.2\), the expected signal amplitude decreases rapidly, and the region of parameter space that can be meaningfully constrained correspondingly narrows and eventually vanishes.} 
All of these exclusion intervals are computed at a \(95\%\) confidence and using the least constraining values for the \ac{BH} age, distance, and mass.


\section{Conclusion}
\label{sec:conclusion}

In this paper, we carry out the first directed searches for long-duration, quasi-monochromatic \acp{GW} from \acp{VBC} around known \acp{BH}. We analyze data from the first part of the \ac{LVK}'s fourth observing run, and we use two semicoherent \ac{CW} search methods, HMM tracking and the Binary BSD-VBC pipeline. Having found no evidence of a \ac{GW} signal, we estimate the range of ultralight vector boson masses that can be constrained. From the HMM search, we disfavor the vector mass ranges $[0.94, 1.08]$ and $[2.75, 3.28] \times 10^{-13}$~eV at 30\% confidence ($P_{\rm fa} = 1\%$). While the present search sensitivity is limited because we target remnant \acp{BH} with SNR values $\lesssim 40$---and thus the confidence level remains statistically insignificant---future searches targeting higher SNR events are expected to yield high-confidence constraints. Meanwhile, the BSD-based search excludes the mass range $[0.85, 1.59] \times  10^{-13}$ eV at \(95\%\) confidence, assuming an initial spin value \(\chi_i > 0.5\) for Cygnus X-1.

As the BSD search demonstrates, we are now able to set constraints on the existence of ultralight vector bosons by targeting known galactic \acp{BH}. Although the constraints obtained from Cygnus X-1 cover only a narrow range of masses, future improvements in detector sensitivity will enhance these constraints. The boson mass ranges that can be constrained also highly depend upon the \ac{BH} parameters. Therefore, running similar searches targeting other known galactic \acp{BH} could extend the exclusion region.

The first HMM search targeting binary merger remnants demonstrates that we are approaching the required sensitivity to place high-confidence constraints on a range of vector masses. Targeting young merger remnants allows us to set independent constraints with minimal assumptions about the \ac{BH}'s history and evolution since formation. Several contributing factors will improve the sensitivity of this type of search: in particular, both increased detector sensitivity and improved search methodologies.
Future observing runs and next-generation \ac{GW} detectors will offer enhanced sensitivity, enabling the detection of numerous binary mergers in the high-SNR regime~\cite{GWTC-4_results, GW_observation_prospects, Evans2021, Punturo2010}. These high-SNR events will yield remnant \acp{BH} with masses and spins that are more accurately and precisely measured. For HMM-based searches, improved posteriors on these parameters will directly translate into tighter constraints on the vector boson mass. 
In addition, future searches can incorporate modifications to the analysis pipelines to further improve search sensitivity (e.g., extending the HMM-based analysis in suitable cases to track time derivatives of the signal frequency within the $\mathcal{F}$-statistic, as demonstrated in Ref.~\cite{Knee2024}).

Observational studies have indirectly constrained the existence of ultralight vector bosons either from \ac{BH} spin measurements~\cite{Baryakhtar2017, Cardoso2018}, by reinterpreting the results of \ac{GW} searches~\cite{Dergachev2019}, or from searches for a stochastic GW background \cite{Tsukada2021}. 
A recent analysis using the GW231123 and GW190517 constituent \acp{BH} disfavors a vector mass range of $[0.11, 18] \times 10^{-13}$~eV assuming a \ac{BH} age of $10^5$~yrs~\cite{aswathi2025}. 
In other searches, the interaction of ultralight vectors with ground-based \ac{GW} detectors has been directly constrained~\cite{direct_dark_matter_LIGO_Virgo, direct_dark_matter_KAGRA}. While each approach involves its own assumptions and limitations, the results presented in this paper provide independent constraints obtained from directed searches that, consistent with previous studies, disfavor the existence of vector bosons with masses of $\sim 1 \times 10^{-13}$~eV.


\section{Acknowledgments}
%
This material is based upon work supported by NSF's LIGO Laboratory, which is a
major facility fully funded by the National Science Foundation.
The authors also gratefully acknowledge the support of
the Science and Technology Facilities Council (STFC) of the
United Kingdom, the Max-Planck-Society (MPS), and the State of
Niedersachsen/Germany for support of the construction of Advanced LIGO 
and construction and operation of the GEO\,600 detector. 
Additional support for Advanced LIGO was provided by the Australian Research Council.
The authors gratefully acknowledge the Italian Istituto Nazionale di Fisica Nucleare (INFN),  
the French Centre National de la Recherche Scientifique (CNRS) and
the Netherlands Organization for Scientific Research (NWO)
for the construction and operation of the Virgo detector
and the creation and support  of the EGO consortium. 
The authors also gratefully acknowledge research support from these agencies as well as by 
the Council of Scientific and Industrial Research of India, 
the Department of Science and Technology, India,
the Science \& Engineering Research Board (SERB), India,
the Ministry of Human Resource Development, India,
the Spanish Agencia Estatal de Investigaci\'on (AEI),
the Spanish Ministerio de Ciencia, Innovaci\'on y Universidades,
the European Union NextGenerationEU/PRTR (PRTR-C17.I1),
the ICSC - CentroNazionale di Ricerca in High Performance Computing, Big Data
and Quantum Computing, funded by the European Union NextGenerationEU,
the Comunitat Auton\`oma de les Illes Balears through the Conselleria d'Educaci\'o i Universitats,
the Conselleria d'Innovaci\'o, Universitats, Ci\`encia i Societat Digital de la Generalitat Valenciana and
the CERCA Programme Generalitat de Catalunya, Spain,
the Polish National Agency for Academic Exchange,
the National Science Centre of Poland and the European Union - European Regional
Development Fund;
the Foundation for Polish Science (FNP),
the Polish Ministry of Science and Higher Education,
the Swiss National Science Foundation (SNSF),
the Russian Science Foundation,
the European Commission,
the European Social Funds (ESF),
the European Regional Development Funds (ERDF),
the Royal Society, 
the Scottish Funding Council, 
the Scottish Universities Physics Alliance, 
the Hungarian Scientific Research Fund (OTKA),
the French Lyon Institute of Origins (LIO),
the Belgian Fonds de la Recherche Scientifique (FRS-FNRS), 
Actions de Recherche Concert\'ees (ARC) and
Fonds Wetenschappelijk Onderzoek Vlaanderen (FWO),
the supercomputing facilities of the Universit\'e catholique de Louvain (CISM/UCL) 
and the Consortium des \'Equipements de Calcul Intensif en F\'ed\'eration Wallonie Bruxelles (C\'ECI),
Belgium,
the Paris \^{I}le-de-France Region, 
the National Research, Development and Innovation Office of Hungary (NKFIH), 
the National Research Foundation of Korea,
the Natural Sciences and Engineering Research Council of Canada (NSERC),
the Canadian Foundation for Innovation (CFI),
the Brazilian Ministry of Science, Technology, and Innovations,
the International Center for Theoretical Physics South American Institute for Fundamental Research (ICTP-SAIFR), 
the Research Grants Council of Hong Kong,
the National Natural Science Foundation of China (NSFC),
the Israel Science Foundation (ISF),
the US-Israel Binational Science Fund (BSF),
the Leverhulme Trust, 
the Research Corporation,
the National Science and Technology Council (NSTC), Taiwan,
the United States Department of Energy,
and
the Kavli Foundation.
The authors gratefully acknowledge the support of the NSF, STFC, INFN and CNRS for provision of computational resources.

This work was supported by MEXT,
the JSPS Leading-edge Research Infrastructure Program,
JSPS Grant-in-Aid for Specially Promoted Research 26000005,
JSPS Grant-in-Aid for Scientific Research on Innovative Areas 2402: 24103006,
24103005, and 2905: JP17H06358, JP17H06361 and JP17H06364,
JSPS Core-to-Core Program A.\ Advanced Research Networks,
JSPS Grants-in-Aid for Scientific Research (S) 17H06133 and 20H05639,
JSPS Grant-in-Aid for Transformative Research Areas (A) 20A203: JP20H05854,
the joint research program of the Institute for Cosmic Ray Research,
University of Tokyo,
the National Research Foundation (NRF),
the Computing Infrastructure Project of the Global Science experimental Data hub
Center (GSDC) at KISTI,
the Korea Astronomy and Space Science Institute (KASI),
the Ministry of Science and ICT (MSIT) in Korea,
Academia Sinica (AS),
the AS Grid Center (ASGC) and the National Science and Technology Council (NSTC)
in Taiwan under grants including the Science Vanguard Research Program,
the Advanced Technology Center (ATC) of NAOJ,
and the Mechanical Engineering Center of KEK.

Additional acknowledgements for support of individual authors may be found in the following document: \\
\texttt{https://dcc.ligo.org/LIGO-M2300033/public}.
For the purpose of open access, the authors have applied a Creative Commons Attribution (CC BY)
license to any Author Accepted Manuscript version arising.
We request that citations to this article use `A. G. Abac {\it et al.} (LIGO-Virgo-KAGRA Collaboration), ...' or similar phrasing, depending on journal convention.




\appendix


\section{Details on BSD search pipeline}
\label{appendix:BSD_method}

\subsection{Signal phase}
\label{appendix:BSD_method_signal}

The phase evolution of a monochromatic signal emitted by a source in a binary system is, in the detector frame, given by
\[\label{eq:phase}
    \Phi(t) = 2\pi f_0 t + \Delta \phi(t),
\]
where \(f_0\) is the signal frequency. The phase modulation due to the combined motion of the source and the detector is given by
\[\label{eq:bsd_modulation}
    \Delta \phi(t) = 2\pi f_0 \qty[ \frac{\vb{r}\vdot\vu{n}}{c} - \frac{R \qty(t + \frac{\vb{r}\vdot\vu{n}}{c} ) }{c} ],
\]
with \(\vb{r}\) the position vector of the detector relative to the \ac{SSB}, and \(\vu n\) the unit vector pointing from the \ac{SSB} to the source. The R{\o}mer delay, \(R\), can be expressed in a low-eccentricity orbit approximation (consistent with the nearly circular orbit of Cygnus X-1), neglecting an irrelevant constant term~\cite{Paola_binary_metric}:
\[
    \begin{split}
        \frac{R(t)}{c} = \ap \bigg[ \sin(\psi(t)) & + \frac{e\cos(\omega)}{2} \sin(2 \psi(t))\\
        & - \frac{e\sin(\omega)}{2} \cos(2 \psi(t)) \bigg],
    \end{split}
\]
where \( \ap \) is the projected semi-major axis, \( e \) is the orbital eccentricity, and \(\omega\) is the argument of periapse. The function \(\psi\) is the mean orbital phase measured from the time of ascending node \( \tasc \), defined as
\[
    \psi\qty(t) = \Omega \qty(t - \tasc),
\] 
with \( \Omega \) the orbital angular frequency.


\subsection{Heterodyne correction}
\label{appendix:BSD_method_heterodyne}

The procedure of the heterodyne correction of \ac{BSD} is presented in Ref.~\cite{Piccinni2019}. It consists of multiplying the data by a complex phase factor \( \exp(-i\,\Delta\phi(t)) \) to correct Doppler-induced variations. The phase \( \Delta\phi(t) \) is computed from Eq.~\eqref{eq:bsd_modulation} assuming a set of parameters \(\vb\Lambda\) to compensate for the delays in the signal arrival time from the motion of the source and the detector. The modulation also depends on the unknown signal frequency. Taking advantage of the \ac{BSD} framework and the downsampling of the data into 1~Hz bands, we substitute for \(f_0\) a reference value fixed at the central frequency. By doing so, we ensure that the error in frequency is bounded by \(\abs{f_0 - f_{0, {\rm ref}}} \leq 0.5\)~Hz.


\subsection{Detection statistic}
\label{appendix:BSD_method_cr}

The \ac{CR} statistic is estimated on a peakmap by projecting the peakmap on the frequency axis and producing the distribution of number of peaks per frequency bin \(n(f)\). We then use a robust estimator defined as~\cite{Pia_method}
\[\label{eq:cr}
    \CR(f) = \frac{n(f) - \bar{n}}{\sigma},
\]
where \( \bar{n} \) is the median of the number of peaks per frequency bin, and the dispersion parameter \( \sigma \) is defined by
\[
    \sigma = \frac{\med{\qty|n(f) - \bar{n}|}}{0.6745}.
\]
The normalization factor \(\sigma\) ensures that if \(n\) follows a normal distribution, then \(\sigma\) is the standard deviation.


\subsection{Template placement}
\label{appendix:BSD_method_grid}

For the placement of templates to cover the \(\ap, \tasc\) parameter space, we use a restricted version of the binary search metric described in Ref.~\cite{Paola_binary_metric}. We use the metric in the semicoherent short-segment regime, where \( \tcoh \ll P \),
\begin{align}
  & g_{\ap\ap} = \frac{1}{6} \qty(\pi\Omega\tcoh f_0)^2,\\
  & g_{\tasc\tasc} = \frac{1}{6} \qty(\pi\Omega^2\tcoh f_0\ap)^2.
\end{align}

The resolution in an orbital parameter \(i\) is then given by~\cite{Paola_binary_metric}\footnote{Note that a factor 2 is missing compared to Ref.~\cite{Paola_binary_metric}. In the present work, these resolutions are used to compute the distance from a template placed on the parameter space rather than the total extent of this template.} 
\[
    \delta_i = \sqrt{0.1 [g^{-1}]^{ii}},
\]
where \(g^{-1}\) is the inverse metric, and the factor 0.1 corresponds to a maximal loss of signal-to-noise ratio of 10\%. Explicitly, we get for the two remaining orbital parameters \(\ap\) and \(\tasc\):
\begin{align}
  & \var\ap = \frac{\sqrt{0.6}}{\pi\Omega \tcoh f_0}, \label{eq:bsd_dap}\\
  & \var\tasc = \frac{\sqrt{0.6}}{\pi\Omega^2 \ap \tcoh f_0}.\label{eq:bsd_dtasc}
\end{align}

Constructing a template grid with varying resolution across the parameter space can be challenging. To simplify this, we adopt a conservative strategy by fixing all the variable parameters to values that maximize grid density. Specifically, we use the maximum frequency \(f_{0,\,{\rm max}}\) within the 1~Hz band, and the maximum value of the projected semi-major axis \(a_{\rm p, max}\) within the range to be covered. Fixing the resolutions as \(\var\ap\qty(f_{0,\,{\rm max}})\) and \(\var\tasc\qty(f_{0,\,{\rm max}}, a_{\rm p, max})\) over the entire parameter space ensures an overcoverage of the search parameter space. With this simplification, the template grid is constructed using a standard square lattice \( \mathbb{Z}_2 \), as described in Refs.~\cite{Paola_binary_metric, Karl_template_placement}.

With this placement strategy, the number of templates needed to cover the parameter space is given by \cite{Paola_binary_metric}
\[\label{eq:grid_size}
    N \approx \left\lceil \frac{\mathcal{R}_{\ap}}{\sqrt 2 \var \ap} \right\rceil  \left\lceil \frac{\mathcal{R}_{\tasc}}{\sqrt 2 \var \tasc} \right\rceil,
\]
where \(\mathcal{R}_{\ap}\) and \(\mathcal{R}_{\tasc}\) are the sizes of the dimensions to be covered.

\subsection{Upper limits formula}
\label{appendix:BSD_ul}

The average prefactor \(\mathcal{B}\) in Eq.~\ref{eq:bsd_sensi} is computed by reproducing the derivation of Eq.(67) in Ref.~\cite{Pia_method} and according to the correction detailed in Ref.~\cite{peakmap_sensi_note}. The general expression of \(\mathcal{B}\) is a function of time, the signal polarization angle \(\psi\), the source sky-position (RA, Dec), and the source inclination angle \(\iota\). It can be expressed as
\[\label{eq:beta_factor}
    \mathcal{B} = \sqrt{\frac{\pi}{2.4308 \left< \qty(F_+ A_+ + F_\times A_\times)^2 \right>_{t}}} \qty(\frac{p_0 (1 - p_0)}{p_1^2})^{1/4}
\]
where the factor \(\frac{\pi}{2.4308}\) is taken from Eq.~B18 in  Ref.~\cite{Pia_method}, and \(p_0\) and \(p_1\) are functions of the peak selection threshold \(\theta_{\rm thr}\)~\cite{Pia_method, peakmap_sensi_note}. For \(\theta_{\rm thr} = 2.5\), we have \(p_0 = 0.075\) and \(p_1 = 0.096\), using the updated definition of \(p_1\) shown in Ref.~\cite{peakmap_sensi_note}. The two beam pattern functions \(F_+\) and \(F_\times\) are defined in Ref.~\cite{JKS}, and the polarization amplitudes are given by \(A_+ = \frac{1 + \cos^2 \iota}{2}\) and \(A_\times = \cos \iota\)~\cite{Pia_method}. 

For the \ac{BSD} search, we evaluate the expression of \(\mathcal{B}\) for the sky-position of Cygnus X-1, and by averaging over the inclination angle error range~\cite{Miller_Jones_CygX1} and the polarization angle \(\psi \in [-\pi/4, \pi/4]\). Using properties of the beam pattern function~\cite{O3_galactic_center}, we can write 
\[
    \begin{split}
        \big< \qty(F_+ A_+ + F_\times A_\times)^2 & \big>_{t, \psi, \cos\iota} \big|_{\rm Dec}\\
        & = \left. \left< F_\times^2 \right>_{t, \psi} \right|_{\rm Dec} \left< A_+^2 + A_\times^2 \right>_{\cos\iota}.
    \end{split}
\]
We compute \( \left< A_+^2 + A_\times^2 \right>_{\cos\iota} = 1.582 \), and, following the development in Ref.~\cite{O3_galactic_center}, \( \left. \left< F_\times^2 \right>_{t, \psi} \right|_{\rm Dec} = 0.197 \) in Livingston and \(0.206\) in Hanford. Injecting these values in Eq.~\ref{eq:beta_factor} gives \(\mathcal{B} \approx 3.37 \) in Livingston and \(\mathcal{B} \approx 3.30\) in Hanford.


\section{Follow-up vetoes}
\label{appendix:follow_up}

As detailed in this section, we only veto candidates that we are confident are caused by noise artifacts. 
The safety of the vetoes used prior to the final manual inspection has been verified in previous analyses employing HMM- and BSD-based techniques, using Monte Carlo simulations in clean frequency bands (see, e.g., Refs.~\cite{ScoX1ViterbiO1, O3_supernova_remnants, O3_galactic_center}).
However, if a \ac{GW} signal is present in the data but overlaps with a noise artifact, it is deemed ``contaminated'' and will be vetoed. In other words, we do not yet have a sufficiently reliable method to separate signals from overlapping, unidentified noise artifacts. Consequently, the false dismissal probability cannot be easily quantified for the vetoes discussed here with the presence of noise artifacts.


\subsection{HMM searches: Merger remnants}
\label{sec:HMM_vetoes}

\subsubsection{Known-line veto}
\label{sec:known_line}

The first veto we use in the HMM search involves comparing the frequency path of each signal candidate against all known instrumental lines present in either the Hanford or Livingston detector to see if there is any overlap~\cite{known_lines_O4a}. We increase the width of the Viterbi path $\delta f \approx 10^{-6}f_i + 8\,\delta f_{\rm SFT}$, where $f_i$ is a given frequency anywhere along the path and $\delta f_{\rm SFT}$ is the SFT frequency bin width. The first part accounts for the Doppler modulation due to Earth's sidereal motion, and the second part accounts for the additional data used in the $\mathcal{F}$-statistic calculation. 
We consider candidates with a wide range of sky positions, start times, and total durations. Because most search timescales are much shorter than a year, we conservatively choose not to factor in Earth's orbital Doppler modulation in this initial veto. The few longer-duration candidates (with $T_{\rm obs} >$ a few months) that could have been vetoed if Earth's orbital Doppler modulation had been included are addressed in later steps.

\subsubsection{Single-interferometer veto}
\label{sec:single_IFO}

Next we use a technique that vetoes candidates caused by noise artifacts in a single detector that are not yet well understood or identified in the official release of \ac{O4a} known instrumental lines. The detailed criteria are as follows:
We run the search with each interferometer individually. Then, a signal candidate can be vetoed as an unknown instrumental line if the detection statistic in one detector is below threshold while the other is greater than the detection statistic from the combined detector search, and if the Viterbi paths of the latter two searches overlap.

\subsubsection{Manual inspection}
\label{sec:manual_inspection}

Because the previous two veto procedures are designed for following up long-duration \ac{CW} search candidates, they are not always able to effectively identify candidates from a short-duration search. Thus, there are many candidates that still remain at this stage. We visually inspect each candidate using a variety of approaches designed to distinguish signals from noise artifacts, typically applying multiple checks to each candidate for confirmation. These approaches are listed here.
\blue{We note that if a true \ac{GW} signal overlaps with a noise artifact, it would be deemed contaminated and vetoed. Since these short-duration searches are particularly susceptible to transient detector noise, injection-based estimation of the false dismissal probability for the manually vetoed candidates is infeasible in practice.}
\begin{enumerate}
\item We relax certain criteria of the single interferometer veto based on close inspection; for example, we veto any candidate whose detection statistic is below threshold in one detector and anomalously high in the other (e.g., $\bar{\mathcal{L}} \gtrsim 100$), regardless of the detection statistic from the original combined search.
\item We compare the frequency drift of the candidate across $T_{\rm obs}$ against the expected frequency drift of a real signal. If the candidate frequency remains within a single bin for at least three quarters of the total duration, \emph{and} if it is not identified by at least one other configuration with a larger value of $T_{\rm coh}$ (which is by design more sensitive to signals with smaller frequency drifts), then it is unlikely to be a real signal. For example, one candidate from GW231123 has an apparent $\dot{f}_0 < 10^{-8}$~Hz\,s$^{-1}$, but the expected frequency drift optimized for $T_{\rm coh} = 14.2$~m is $\dot{f}_0 \sim 7 \times 10^{-7} $~Hz\,s$^{-1}$.
This veto criterion is typically applied in conjunction with at least one other veto as a cross-check.
\item We assess whether the detection statistic of a given candidate is consistent with expectations for a signal from the remnant \ac{BH} given its distance estimate. If the statistic is significantly larger than expected (e.g., $\bar{\mathcal{L}} \gtrsim 100$), we can veto the candidate as a loud but unidentified noise artifact. This veto is applied only in conjunction with at least one other veto. For example, if the detection statistic combining two detectors is significantly larger than expected for the target's distance, it is usually accompanied by the candidate being much louder in one detector than in the other, in which case the candidate is also vetoed according to the first manual inspection criterion described above.
\item We examine the signal candidate in the spectrogram from each interferometer in the relevant frequency band and time segment to determine whether it overlaps with any visible noise artifacts.
\item Finally, if two candidates with the same frequency path are found using two different search configurations---in other words, if the same candidate is identified in two different searches---and if one of them has been vetoed by any of the above criteria, we consider the other likely to have arisen from the same artifacts. This veto criterion is typically applied in conjunction with other manual inspections as a cross-check.
\end{enumerate}


\subsection{BSD search: Cygnus X-1}
\label{appendix:BSD_follow_up}

For the BSD search, the 27 stage-2 candidates have been filtered by removing those associated with known instrumental lines~\cite{known_lines_O4a} and clustering candidates that shared a common candidate. Six candidates remained after this filtering and are listed in Table~\ref{tab:BSD_candidates}. The values reported in the table correspond to the averages of the template parameters between the two detectors, and the quoted uncertainties are obtained by propagating the template resolutions given in Eqs.~\ref{eq:bsd_dap} and~\ref{eq:bsd_dtasc}. As discussed in Sec.~\ref{sec:bsd_grid}, the other parameters, \(\{{\rm RA}, {\rm Dec}, P, e, \omega\}\), are fixed to the central values listed in Tables~\ref{tab:BH_params} and~\ref{tab:cygx1-para}, and are therefore identical for all the candidates.

Given the small number of surviving candidates, we conducted the following tests and visually inspected the results for each. Although we did not formally estimate the false dismissal probability, the validity and safety of each veto have been confirmed by applying the follow-up procedure to simulated signals with parameters similar to those of the six candidates. The tests, adapted from previous \ac{BSD}-based searches (e.g.,~\cite{O3_galactic_center, O3_supernova_remnants}), are grouped into three categories: (i) consistency of the candidate with its astrophysical parameters, (ii) consistency of the detection statistic with different search configurations, and (iii) a manual inspection where we associated some candidates to artifacts caused by non-Gaussianities in the detector. These tests are described in detail below.

\begin{table}[tbh]
    \centering
    \setlength{\tabcolsep}{8pt}
    \renewcommand\arraystretch{1.2}
    \caption{Parameters of the candidates from the BSD search, averaged between the two detectors. For each candidate, we indicate the test by which it was vetoed.}
    \begin{tabular}{cccc}
        \hline
        \hline
        \(f_0\) [Hz] & \(\ap\) [s] & \(\tasc\) [s] & Veto \\
        \hline
        32.751 & \(38.84 \pm 0.82\) & \(5\,600 \pm 1\,700\)     & Stat. \\
        37.131 & \(36.09 \pm 0.73\) & \(188\,100 \pm 1\,600\)   & Manual \\
        41.176 & \(36.79 \pm 0.66\) & \(-4\,383 \pm 1\,400\)    & Astro. \\
        41.627 & \(38.60 \pm 0.65\) & \(-11\,800 \pm 1\,300\)   & Stat. \\
        53.890 & \(40.62 \pm 0.50\) & \(204\,700 \pm 950\)      & Manual \\
        96.114 & \(38.10 \pm 0.28\) & \(223\,790 \pm 570\)      & Manual \\
        \hline
        \hline
    \end{tabular}
    \label{tab:BSD_candidates}
\end{table}

\subsubsection{Known-line veto}

Similar to the HMM method, we veto any candidate whose frequency evolution crosses a known instrumental line in either the Hanford or Livingston detector~\cite{known_lines_O4a}. For a candidate observed at a frequency \(f_0\), we estimate the modulation range as \(f_0 \pm \delta f\), where \(\delta f\) is the modulation size computed with the parameters of the candidate,
\(
    \delta f = \max_t\qty(\frac{\rm{d} \Delta \phi(t)}{{\rm d} t}),
\)
where \(\Delta \phi(t)\) is the phase modulation defined in Eq.~\ref{eq:bsd_modulation}.

\subsubsection{Clustering}

During the coincident candidate selection process, we consider all possible pairs of candidates from both detectors. Some redundancy is therefore possible in the list of candidates. Indeed, from the eight candidates passing the known-line veto, two candidates are repeated twice, in association with two different (but similar) candidates in the second detectors. We therefore cluster these two sets of candidates and follow them up conjointly. The number of unique candidates is reduced to six.

\subsubsection{Astrophysical consistency veto}

The second step of the follow-up procedure involves assessing the consistency of each candidate with an astrophysical signal through two key checks. Any candidate failing at least one of these tests is discarded from further analysis.

\begin{enumerate} 
\item \textbf{Frequency refinement.}
Since the search demodulates the data using the central frequency of each 1~Hz band, a real signal may not be perfectly corrected if its true frequency differs from this reference.
To refine the analysis, we re-run a localized search using the candidate recovered frequency as the reference frequency for demodulation. A genuine signal must persist with similar or improved significance under this correction.

\item \textbf{Unmodulated test.}
As the search is directed toward a signal emitted from Cygnus X-1, the significance of an astrophysical signal should decrease when using parameters incompatible with the target. In particular, we try to recover the candidate without performing the heterodyne correction. A candidate persisting with comparable significance in this configuration cannot be of astrophysical origin and is therefore vetoed.
\end{enumerate}


\subsubsection{Statistical consistency check}

\begin{enumerate} 
\item \textbf{Sensitivity vetoes.}
For a true signal, the detection statistic should scale with the sensitivity of each detector. We check this by normalizing the \ac{CR} with the median \ac{ASD} of the 1~Hz band, \( \sqrt{S_n} \). We then require
\[
    \frac{\CR_1}{\sqrt{S_{n_1}}} < 3 \frac{\CR_2}{\sqrt{S_{n_2}}},
\]
where detector 1 is less sensitive and detector 2 is more sensitive. The factor 3 is a conservative choice and indeed none of the candidate was vetoed by this check.

\item \textbf{Cumulative and Uniformity veto.}
A genuine astrophysical \ac{CW} signal should persist throughout the observation run, and we expect: (i) steadily increasing significance as more data are included, and (ii) the candidate to be present in any subset of the data.

We first examined cumulative behavior by computing the \ac{CR} and signal-to-noise ratio (SNR) over data segments whose durations increased in 30-day steps, for both detectors and for data with and without heterodyne correction. We compared the corrected and uncorrected results, looking for behavior inconsistent with the presence of a signal. One candidate was vetoed because the uncorrected data yielded higher significance for most of the run. A second candidate showed a sharp increase during the first month, indicating the presence of non-stationary noise.

To verify the persistence across subsets, we analyzed one-month segments with various start times, again for both detectors and both data types. Some variation is expected due to varying duty cycles or noise levels, so results were interpreted with tolerance. For the second suspect candidate identified in the cumulative test, corrected and uncorrected data gave similar \ac{CR} values except in the first month, where the corrected data produced a much higher \ac{CR}. This confirmed the non-persistent nature of the candidate, and it was vetoed.
\end{enumerate}

\subsubsection{Manual inspection}
Finally, we manually inspect the three remaining candidates. For each, we investigate the spectra of the corrected and uncorrected data using two different frequency resolutions: \(5.5\times 10^{-4}\) and \(1.1\times 10^{-5}\)~Hz, corresponding to coherence times of 30~min and 1~day, respectively. In all three candidates' bands, we observe strong non-Gaussian noise profiles. Such non-Gaussianities are known to produce artifacts in specific frequency bins during peakmap peak selection~\cite{Peakmap_method}, and, coincidentally, all the remaining candidates are present in one of these frequency bins. By slightly modifying the resolution of the background estimation used to compute the equalized spectra~\cite{Peakmap_method}, we force the artifact-affected bins to be moved away from the candidate bins. This modified setup ensures that the candidates' frequency bins were not contaminated by the peakmap creation artifacts. We tested several such configurations, and in all cases the \ac{CR} of each candidate dropped well below the selection threshold.
\blue{For each of the three candidates, we have assessed the impact of these modified configurations on a simulated signal injection. The injections were done with the same orbital parameters as the candidates. We have chosen the injection amplitudes from Eq.~\eqref{eq:bsd_sensi} by replacing \({\rm CR}_{\rm max}\) with the threshold value in band as defined in Sec.~\ref{sec:bsd_cand_selection}. We observed that all the injections remained detectable in all configurations considered. On this basis, and given that candidates are not significant, in the sense discussed in Sec.~\ref{sec:bsd_cand_selection}, in unaffected configurations, we vetoed all remaining candidates.}

\def\bibsection{\section*{References}}
\bibliography{O4_vector_bosons}

\end{document}